\documentclass[reprint,aps,prd,]{revtex4-1}
\usepackage{dcolumn}
\usepackage{amsmath}
\usepackage{amssymb}
\usepackage{bm}
\usepackage{booktabs}
\usepackage{graphicx}
\usepackage{subfig}

\usepackage{multirow}

\usepackage{hyperref}
\hypersetup{colorlinks=true, allcolors=blue} 

\usepackage{color}
\usepackage{ulem}
\usepackage{marginnote}

\usepackage[T1]{fontenc}
\usepackage[utf8]{inputenc}
\usepackage{textcase}
\usepackage{geometry}
\usepackage[table]{xcolor}
\usepackage{textcomp}
\usepackage{siunitx}

\usepackage{hyperref}
\hypersetup{
    colorlinks=true,
    linkcolor=blue,
    filecolor=magenta,      
    urlcolor=cyan,
    citecolor=blue,
}

\begin{document}

\title{Experimental Realization of Optimized Ternary Mirror Coatings}

\author{V. Pierro,$^{1,2}$ M. Granata,$^{3}$ C. Michel,$^{3}$   L. Pinard,$^{3}$ B. Sassolas,$^{3}$ D. Forest,$^{3}$ N. Demos,$^{4}$ S. Gras,$^{4}$ M. Evans,$^{4}$ I. M. Pinto,$^{2}$ G. Avallone,$^{5}$ V. Granata,$^{6,2}$ }

\affiliation{$^{1}$Dipartimento di Ingegneria (DING), Universit\`a del Sannio, Piazza Roma, 28, 82100 Benevento BN, Italy}
\affiliation{$^{2}$INFN Sezione di Napoli Gruppo Collegato di Salerno, Complesso Universitario di Monte S. Angelo, 80126 Napoli, Italy} 
\affiliation{$^{3}$Laboratoire des Mat\'{e}riaux Avanc\'{e}s - IP2I, CNRS,  Universit\'{e} Claude Bernard Lyon 1, 
F-69100 Villeurbanne, France}
\affiliation{$^{4}$Massachusetts Institute of Technology, 185 Albany Street NW22-295, Cambridge, Massachusetts 02139, USA}
\affiliation{$^{5}$ Dipartimento di Fisica “E.R. Caianiello”, Università di Salerno, Via Giovanni Paolo II, 132, 84084 Fisciano SA, Italy}
\affiliation{$^{6}$ Dipartimento di Ingegneria Industriale, Elettronica e Meccanica DIIEM, Università di Roma Tre, Via Vito Volterra, 62, 00146 Roma RM, Italy}


\begin{abstract}
We report on the first experimental realization of multi-material dielectric mirror coatings designed through a multi-objective optimization algorithm to simultaneously minimize thermal noise and optical losses.
 We validate this design strategy by fabricating and characterizing two distinct ternary systems: a SiN\(_x\)-based proof-of-concept and a Ti:GeO\(_2\)-based system targeting lower optical losses. 
 The performance of the SiN\(_x\) coating shows remarkable agreement with predictions, demonstrating a noise amplitude spectral density reduction of 0.82 with respect to current reference coatings, and validating our design-to-fabrication pipeline.
  The Ti:GeO\(_2\)-based system achieves the crucial goal of sub-ppm absorption; its measured thermal noise, however, is higher than the theoretically predicted level of 0.71, extrapolated from single-layer material characterization. 
  A dedicated tolerance analysis confirms that this discrepancy is not attributable to random thickness errors, emphasizing that further studies of the manufacturing process are needed to fully exploit this combination of materials. 
  This work establishes a robust methodology for producing complex, high-performance optical coatings tailored for  precision experiments.
\end{abstract}

\maketitle

{\it Introduction$-$} 
In the last two decades, optics has reached unprecedented levels of precision and stability. In recognition of this, the United Nations declared 2022 the International Year of Glass, following the International Year of Light, which commemorated the 1,000th anniversary of Al-Hasan Ibn Al-Haytham’s \textit{Kitab al-Manazir} (Book of Optics) \cite{IYL}.

For many advanced optical applications, including high-power \cite{jawad24,turnbaugh21,zhu18} and ultra-stable lasers \cite{USL2}, optical cavities for precision experiments \cite{CPM1}—even in fundamental physics \cite{PVLAS,granata20_review,GW}—and photonic devices \cite{fotonica}, the reduction of Brownian thermal noise remains a central challenge despite these advances.

A major obstacle in designing coatings for high-sensitivity optical devices is the difficulty of identifying materials that simultaneously provide low optical absorption, low Brownian noise, and a sufficiently broad range of refractive indices.

A promising approach is the use of multi-material coatings, which allow thin low-noise but absorptive layers to be buried under noisier yet low-absorption layers \cite{yam15,steinlechner15, Demos2025}. 
Theoretical formulations of this design problem \cite{VPbinary,BORGVP}, as well as initial demonstrations of feasibility \cite{pierro21}, have been reported. 
Multi-material coatings have been proposed for gravitational-wave interferometers \cite{MMforET}, and some preliminary experimental realizations have been achieved as proof-of-concept \cite{craig20,MMchao}.

In this work, we take a decisive step beyond previous empirical demonstrations: for the first time, we realize two distinct multi-material coatings, both designed through a fully predictive optimization framework. 
This achievement demonstrates the feasibility of producing advanced ternary coatings by means of an innovative design-to-fabrication pipeline, establishing a new paradigm for the development of high-performance optical coatings.

{\it Design Strategy $-$} 
To overcome the performance limits of traditional binary stacks, we developed a design framework for multi-material coatings based on a Double-Stack of Doublets (DSD) architecture (see  {\it supplementary material} \cite{supp} for details). 
This design consists of two sub-stacks: a bottom stack near the substrate, made of materials with high refractive index contrast (like SiN$_x$/SiO$_2$ or Ti:GeO$_2$/SiO$_2$ ) 
to achieve high reflectivity with fewer layers, and a top stack, made of materials with low optical absorption (Ta$_2$O$_5$-TiO$_2$/SiO$_2$), to shield the more absorptive layers from the incident laser field. 
The layer thicknesses are not quarter-wave but are  determined by a multi-objective evolutionary algorithm \cite{hadka13} that seeks Pareto-optimal (best-tradeoff)  solutions 
for thermal noise, absorption, and transmittance \cite{pierro21}.

Our methodology avoids subjective weighting of the objectives and employs a deterministic, two-stage selection process, as detailed in the supplementary material \cite{supp},
which leads to the identification of a single optimal design.

To compute the coating properties, our algorithm used a transmission matrix method for the optics \cite{abele50} and a thermal noise model by Braginsky et al. \cite{braginsky12}. 
In this model, the noise power spectral density $S_{CB}$ is a linear combination of the contributions from each layer, weighted by its thickness $d_i$ and its coefficient $\eta_i$:
\begin{equation}
S_{CB}(f) = \frac{2 k_B T}{\pi^2 w^2 f} \sum_{i = 1}^{N} \eta_i d_i,
\label{eq:noise}
\end{equation}
where
\begin{equation}
\eta_i = \varphi_i  \left[ \frac{(1+\sigma_i)(1-2\sigma_i)}{Y_i(1-\sigma_i)} + \frac{Y_i(1+\sigma)^2(1-2\sigma)^2}{Y^2(1-\sigma_i^2)} \right].
\label{EQ_BGV_coeff}
\end{equation}
In Eq. (\ref{EQ_BGV_coeff}), $k_B$ is the Boltzmann constant, $T$ is the mirror temperature, $w$ is the beam radius, $f$ is the frequency, $Y$ and $\sigma$ are the Young modulus 
and Poisson coefficient of the substrate, respectively, and $d_i$, $Y_i$, $\sigma_i$ and $\varphi_i$ are the thickness, Young modulus, Poisson coefficient and the elastic loss angle of the $i$-th layer, respectively. 
The loss angle $\varphi_i$ quantifies the layer elastic energy dissipation and determines its noise amplitude \cite{callen52}. 
Knowledge of the $\eta_i$ coefficients is thus crucial for making reliable noise predictions. A detailed discussion of the noise model and its limitations is provided in the supplementary material \cite{supp}.

In this work, we used silicon-rich silicon nitride (SiN$_x$) and Ti-doped germania (i.e. Ti:GeO$_2$), as layers of high refractive index, lower noise and higher absorption \cite{granata20_review}.
The other materials used were Ti-doped tantala (i.e. Ti:Ta$_2$O$_5$) as high-index layers of higher noise and lower-absorption, 
and silica (SiO$_2$) as low-index layers of low noise and absorption, as in current mirror coatings of interferometric gravitational wave detectors \cite{granata20}.

The use of the DSD structure is certainly motivated by its simplicity (theoretical and experimental), but also by certain incompatibilities between the different synthesis conditions of the materials.
In particular, a limitation is posed by the onset of crystallization in the Ti:Ta$_2$O$_5$ layers upon thermal treatment ({\it annealing}), which is standard practice.
While the SiN$_x$ and SiO$_2$ layers might be annealed up to $1000$ $^{\circ}$C without crystallization \cite{MGr,silenzi25}, the amorphous to poly-crystalline phase transition of the tantala-titania layers 
takes place between $650$ and $700$ $^{\circ}$C. 
To take advantage of the low loss angle of high-temperature annealed SiN$_x$ and SiO$_2$, we split our design into two distinct doublet-based binary stacks.

In line with the stringent requirements for mirror coatings in gravitational-wave interferometers, which often include constraints at auxiliary 
wavelengths (e.g., $532$\,nm for thermal control systems), our design prioritizes the primary $1064$\,nm detection band.

We focus on a {\it single-band} optimization because a {\it dual-band} approach, which our code can readily handle, would compromise performance in the detection band,
our primary goal being to establish the ultimate performance limit, in terms of thermal noise, at the operating wavelength (see rationale in \cite{supp}).
The main constraint is therefore a maximum transmittance ($\tau_c$) of $5.6$\,ppm at a $1064$\,nm wavelength.
The significant difference in the extinction coefficients of silicon nitride (SiN$_x$) and titanium-doped germanium dioxide (Ti:GeO$_2$), 
however, directly impacts the fundamental absorption limits \cite{supp} of their DSD designs. This is clearly observed in simpler binary systems, where the minimum achievable absorption for SiO$_2$/Ti:GeO$_2$ designs is fundamentally 
lower than for SiO$_2$/SiN$_x$ designs. 
To account for this difference, we have set different absorption targets:  $0.5$\,ppm for the Ti:GeO$_2$ design and $1.5$\,ppm for the SiN$_x$ design. 

\begin{figure}[h!]
	\centering
    \subfloat[SiN$_x$-based design]{\includegraphics[width=0.95\columnwidth]{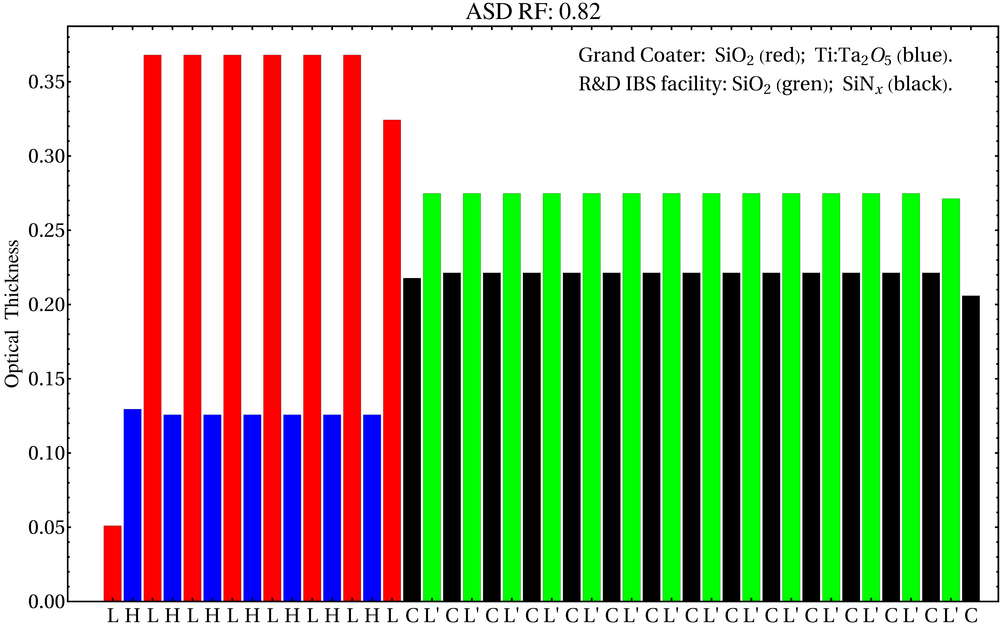}\label{FIG_design_sinx}}
    \hfill
    \subfloat[Ti:GeO$_2$-based design]{\includegraphics[width=0.95\columnwidth]{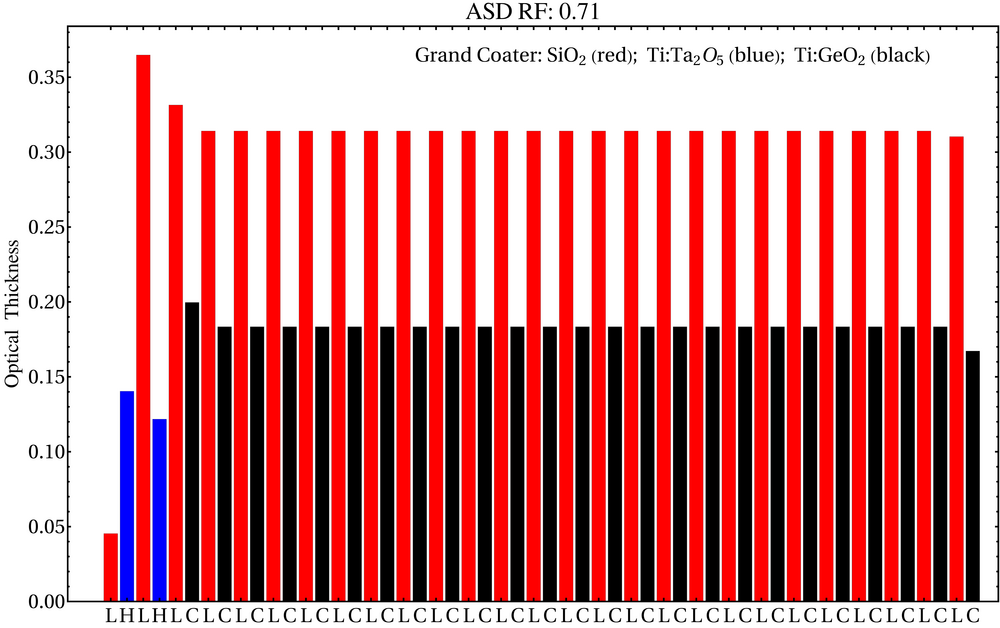}\label{FIG_design_geo2}}
	\caption{
	Layout and predicted thermal noise performance (ASD RF) of the two optimized ternary mirror coatings realized in this work.
	Vacuum is on the left (the top) of the design, substrate is on the right of the figure. 
	 The bar colors correspond to the different dielectric materials. 
	 For the SiN$_x$-based design (a), the bottom stack consists of SiN$_x$ (black) and SiO$_2$ (green) 
	 both deposited in an {\it R\&D} IBS facility at LMA. The top stack consists of 
	 Ti:Ta$_2$O$_5$ (blue) and SiO$_2$ (red), both deposited in the main IBS facility at LMA (the Grand Coater). 
	 For the Ti:GeO$_2$-based design (b), the bottom stack uses Ti:GeO$_2$ (black) and SiO$_2$ (red), 
	 while the top stack uses Ta$_2$O$_5$-TiO$_2$ (blue) and SiO$_2$ (red), all deposited in the LMA Grand Coater.}
\label{fig:designs}
\end{figure}

The optimized DSD coating designs are presented in Fig.~\ref{fig:designs}{, where layer thicknesses are plotted in optical units.
The material refractive indices required to convert these to physical thicknesses are listed in Table \ref{tab:tab_refl}.}
The accompanying {\it supplementary material} \cite{supp} provides a comprehensive description of the optimization framework, the resulting designs'  full optical characteristics, 
and a crucial baseline analysis. This baseline explains why these materials underperform in traditional binary stacks, thereby motivating the necessity of the DSD architecture developed here.
\reversemarginpar
Table \ref{tab:tab_optical} lists the predicted optical properties of the final designs. 
While the transmittance and absorption at $1064$\,nm adhere to the predefined design constraints, 
the transmittance at $532$\,nm is a resultant property of the {\it single-band} optimization.
The reflection coefficient at the first interface is close to $-1$, 
indicating a near-zero electric field at the surface, which minimizes sensitivity to surface contamination.

{\it Experimental Validation and Results $-$} 
To validate our design-to-fabrication pipeline, we realized the two ternary DSD coatings through a multi-step process, 
followed by a comprehensive and thorough campaign of characterization  of their optical and mechanical performances.

The multi-material coating samples were grown at the Laboratoire des Mat\'eriaux Avanc\'es \cite{LMA}
using a two-step deposition process that, for the SiN$_x$ case, required two different custom-developed ion-beam sputtering (IBS) systems. 
The lower SiN$_x$/SiO$_2$ stack was first grown in the so-called {\it R\&D} system, 
after which the samples were transferred to the so-called {\it Grand Coater} for the deposition of the upper Ta$_2$O$_5$-TiO$_2$/SiO$_2$ stack. 
The use of two distinct systems was mandated by scheduling constraints on the Grand Coater, 
which is the primary machine for manufacturing mirror coatings for current gravitational-wave interferometers \cite{granata20}.
In both systems, samples were grown using accelerated Ar ions ($\sim$1 keV, $\sim$1 A) as sputtering particles. 
The working pressure during deposition was maintained at $\sim 10^{-4}$\,mbar, with O$_2$ fed into the chamber for oxide layers \cite{granata20,MGr}.

Transmittance spectra were acquired over the $400$–$1400$\,nm range using a PerkinElmer Lambda 1050 spectrophotometer.
Optical absorption at $1064$\,nm was measured with a precision of $<0.5$\,ppm using a custom setup based on 
photo-thermal deflection \cite{Boccara80}. 
The critical Brownian thermal noise was measured in the $30$\,Hz to $3$\,kHz frequency band using a folded Fabry-Perot cavity  \cite{Gras18,Gras17}. 
 This technique employs co-resonating TEM$_{02}$ and TEM$_{20}$ orthogonal laser modes to precisely isolate the coating's noise spectrum. 
 The optical and mechanical parameters (refractive index, absorption, and internal friction) of the individual materials used in the design 
 algorithm were sourced from prior, dedicated characterizations of single-layer samples \cite{granata20_review,granata20,Granata21}.

{\it SiN$_x$-based Coating.} 
The first design uses a bottom stack of high contrast SiN$_x$/SiO$_2$ and a top stack of low-absorption Ti:Ta$_2$O$_5$. 
The optimized structure is shown in Fig. \ref{fig:designs}a.
A split-annealing process was required to prevent the onset of crystallization in Ta$_2$O$_5$-TiO$_2$ . 
We separately annealed the lower SiN$_x$/SiO$_2$ stack at higher temperature (900 $^{\circ}$C for 10 hours), before depositing  the upper Ta$_2$O$_5$-TiO$_2$/SiO$_2$ stack and 
annealing the whole DSD design at a lower temperature (600 $^{\circ}$C for 10 hours). 
The measured and theoretical transmittance spectra for the SiN$_x$-based design is displayed in Fig. \ref{fig:Spectrum}a,  the experimental spectrum shows remarkable agreement with the 
theoretical curve, accurately reproducing the central wavelength and width of the high-reflectivity band, as well as the pass-band ripples.
The modest accuracy at shorter wavelengths, such as $532$\, nm, was anticipated, since the theoretical spectrum was generated using constant 
refractive indices (fixed at their $1064$\, nm values), thereby not accounting for the effects of material dispersion.

\begin{figure}[h!]
\centering
    \subfloat[SiN$_x$-based design]{\includegraphics[width=0.95\columnwidth]{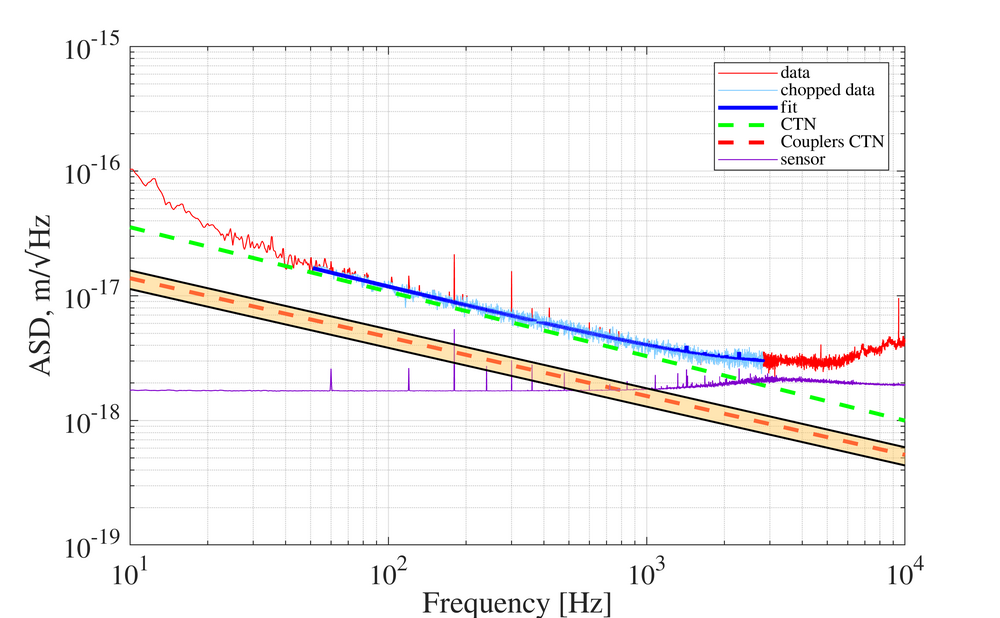}\label{FIG_noise_sinx}}
    \hfill
    \subfloat[Ti:GeO$_2$-based design]{\includegraphics[width=0.95\columnwidth]{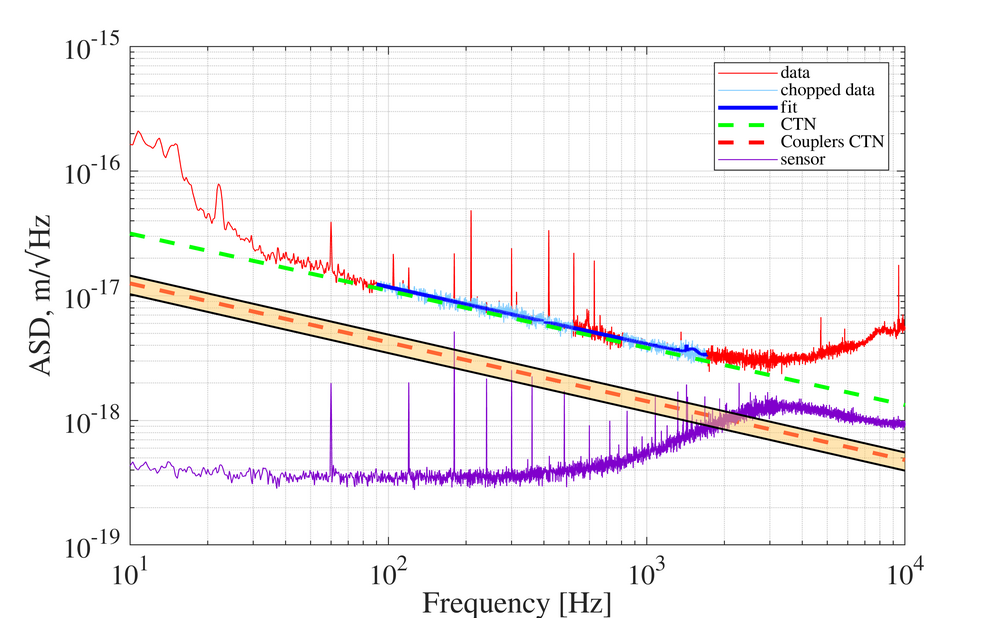}\label{FIG_noise_geo2}}
\caption{CTN measurements of the two optimized ternary mirror coatings. (a) SiN$_x$-based design. (b) Ti:GeO$_2$-based design. 
	The experimental noise is presented as both a raw measurement ({\it data}) and a filtered trace ({\it chopped data}), from which noise around the harmonics of the 60 Hz power-line 
	frequency has been removed. 
	The solid curve, labeled {\it fit}, represents the estimated thermal noise.}
\label{fig:noise}
\end{figure}

Figure~\ref{fig:noise}a presents the measured amplitude spectral density (ASD) for 
the optimized SiN$_x$-based design,  obtained from noise measurements performed with the Coating Thermal Noise (CTN) interferometer. Again, the observed performance 
 shows remarkable agreement with predictions (the measured properties in Table~\ref{tab:results} closely match the design targets in Table~\ref{tab:tab_optical}), 
exhibiting a thermal noise of $10.6 \times 10^{-18}$~m/$\sqrt{\text{Hz}}$. 
This corresponds to an ASD Reduction Factor (ASD RF) of 0.82 
relative to the current LIGO and Virgo End Test Mass (ETM) coatings~\cite{MGr}. 
Furthermore, the measured spectral index of $\alpha = 0.5 \pm 0.03$ indicates 
a negligible dispersion of the mechanical parameters.

In addition, an absorption of approximately $1.5$~ppm was measured. 

Collectively, these results provide the crucial experimental validation of our entire methodology.

{\it Ti:GeO$_2$-based Coating.} 
To demonstrate the full potential of our validated framework, we applied it to a more advanced material system. 
We replaced the SiN$_x$/SiO$_2$ bottom stack with  a Ti:GeO$_2$/SiO$_2$ one, a combination known for its potential for lower intrinsic losses \cite{supp}. 
The resulting optimized design is shown in Fig. \ref{fig:designs}b
and its transmittance spectrum is shown in Fig. \ref{fig:Spectrum}b showing a fair match between measurement and prediction.

The final absorption of $0.7$\, ppm (see Table \ref{tab:results}) was achieved through a dedicated multi-step procedure:  
first, the substrate underwent a $10$-hour pre-annealing at 900$^{\circ}$C. 
Subsequently, after the deposition of the first layer stack, a $100$-hour anneal at $600$ $^{\circ}$C was performed.
Finally, after the second deposition step, the entire coating was subjected to a final $100$-hour annealing at 500$^{\circ}$C.

As detailed in Table \ref{tab:results}, the resulting mirror exhibits good performance, with a measured thermal noise of
 $11.2 \pm 0.2 \times 10^{-18}$ m/$\sqrt{\text{Hz}}$ , corresponding to an ASD RF of $0.81$.
The spectral index of $\alpha = 0.44$ confirms the presence of mechanical dispersion. 

\begin{figure}[h!]
\centering
    \subfloat[SiN$_x$-based design]{\includegraphics[width=0.95\columnwidth]{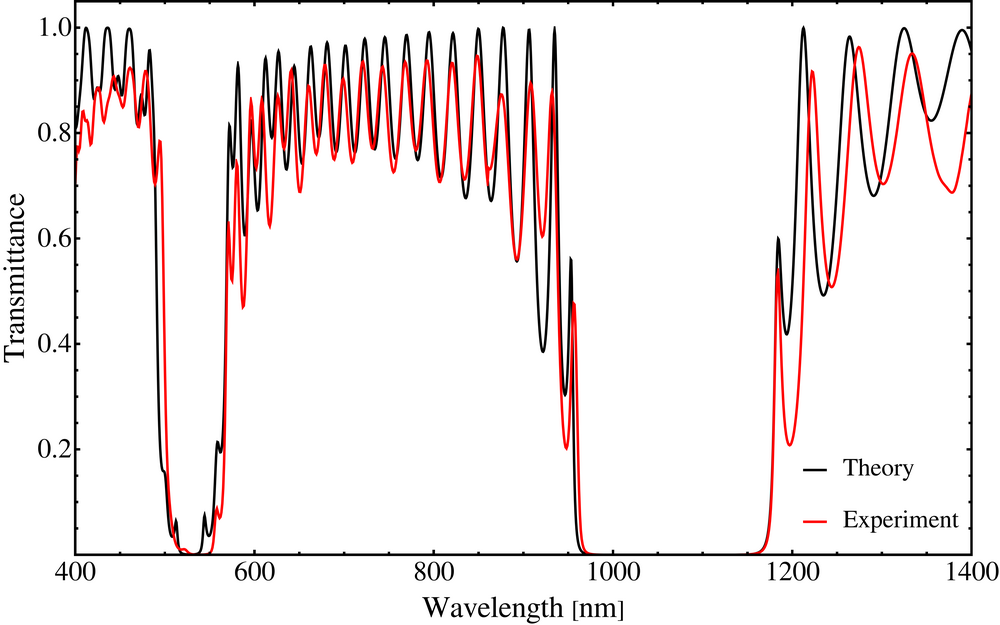}\label{FIG_spectr_sinx}}
    \hfill
    \subfloat[Ti:GeO$_2$-based design]{\includegraphics[width=0.95\columnwidth]{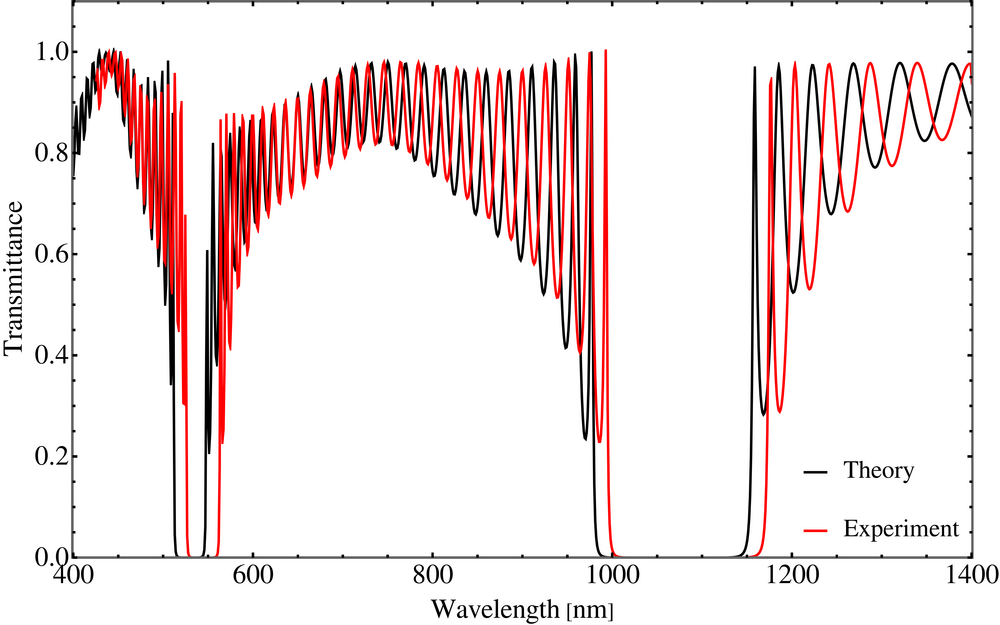}\label{FIG_spectr_geo2}}
	\caption{Comparison of the measured (red curve) and theoretical (black curve) transmittance spectra for the two optimized ternary coatings: 
	(a) the SiN$_x$-based design and (b) the Ti:GeO$_2$-based design. 
	For both coatings, an excellent agreement is observed around the $1064$ nm design wavelength, especially for the SiN$_x$ based structure, 
	validating the {\it quasi}-periodic design and the high precision of the fabrication process. 
	The minor deviation at shorter wavelengths is expected, as the theoretical model does not account for material dispersion, assuming instead constant 
	refractive indices based on their values at 1064 nm.
}
\label{fig:Spectrum}
\end{figure}

While the coating successfully achieved the goal of sub-ppm absorption (below its Carniglia limit \cite{supp}), 
the measured  thermal noise performance (ASD RF of $0.81$) did not fully meet the ambitious design target (ASD RF
of $0.71$ \cite{supp}). This deviation, mirrored by a measured transmittance of $8.8$ ppm which exceeds the $5.6$ ppm design constraint, along with 
the less perfect agreement in the transmittance spectrum (Fig.~\ref{fig:Spectrum}b), suggests 
that realizing the full potential of this advanced material combination requires further study of the fabrication process,
and its impact on the relevant material properties.
To further investigate the source of this deviation, a dedicated tolerance analysis was performed in \cite{supp}. 
The simulations show that the design is robust against independent

random layer thickness errors ($\pm 0.5$\, nm for each layer) {as well as against uncertainties in the material refractive indices (0.2\%)},
 with the ASD RF exhibiting a nearly Gaussian distribution with a standard deviation of less than 0.01\%. 
Such a minimal variation from fabrication tolerances cannot account for the significant gap between the predicted and measured performance. 
This strongly suggests that the discrepancy is not due to simple stochastic errors but
points to a more fundamental issue, such as an unforeseen material incompatibility or process-induced stress that is
not captured by single-layer characterization. This confirms a crucial point: while the choice of materials sets
the performance limit, mastering the complex interplay between materials during co-deposition and annealing is
paramount to fully exploiting their potential.

\begin{table*}[h!]
\centering
\caption{Real part ($n$) and extinction coefficient ($k$) of the complex refractive index for the coating materials used in this work, 
specified at the 1064\, nm design wavelength. Values are sourced from prior characterizations: 
SiO\textsubscript{2} and Ti:Ta\textsubscript{2}O\textsubscript{5} from \cite{granata20}, SiN\textsubscript{x} from \cite{MGr}, and Ti:GeO\textsubscript{2}
 is consistent with data in \cite{granata20_review}.}
\label{tab:tab_refl}
\begin{tabular}{|l|c|c|c|}
\hline
\hline
Material & Deposition System & Real Part ($n$) & Extinction Coeff. ($k$) \\
\hline
SiN\textsubscript{x}                & R\&D           & 2.05 & $1.5 \times 10^{-5}$ \\ \hline
SiO\textsubscript{2}                 & R\&D           & 1.475 & $5 \times 10^{-8}$ \\ \hline
SiO\textsubscript{2}                 & Grand Coater   & 1.45 & $3 \times 10^{-8}$ \\ \hline
Ti:Ta\textsubscript{2}O\textsubscript{5} & Grand Coater   & 2.09 & $5 \times 10^{-8}$ \\ \hline
Ti:GeO\textsubscript{2}              & Grand Coater   & 1.89 & $1.9 \times 10^{-7}$ \\ 
\hline
\hline
\end{tabular}
\end{table*}

\begin{table*}[h!]
\caption{ Optical properties computed via optimization for both SiN$_x$ and Ti:GeO$_2$ based designs.}
\label{tab:tab_optical}
\begin{tabular}{|l|l|c|c|c|c|}
\hline
\hline
DSD design & Complex reflection  & Transmittance & Absorption & Transmittance  & ASD RF\\ 
   &  coefficient at vacuum interface & @ 1064\, nm & @ 1064\, nm & @ 532\, nm &      \\  \hline
SiN$_x$-based  & $-0.99814 - \imath \, 0.060886$ & $5.6$\, ppm & $1.45$\, ppm & $11.6$\,  ppm & 0.82\\ \hline
Ti:GeO$_2$-based  & $-0.998396 - \imath 0.0565713 $ & $5.6$\, ppm & $0.5$\, ppm &  $55.25$ \, ppm& 0.71\\ \hline
\hline
\end{tabular}
\end{table*}

\begin{table*}[h!]
\centering
\caption{Measured properties of the realized ternary mirror coatings compared to current technology.}
\begin{ruledtabular}
\begin{tabular}{|l|ccc|cc |c|}
System & Deposition &  Absorption [ppm] & Transmittance [ppm] & \multicolumn{2}{c}{$S_{CB}(f) = A \times (100\ \text{Hz}/f)^{\alpha}$} & ASD RF\\
                 &                   &           &       & A $(10^{-18} \text{m}/\sqrt{\text{Hz}})$ & $\alpha$  &   \\
\colrule
Current ETM &  Standard & 0.5 $\pm$ 0.2 & 6.0 $\pm$ 0.2& 13.7 $\pm$ 0.3 & $0.45 \pm 0.02$ & 1\\
\hline
SiN$_x$-based DSD &  Split & 1.4 $\pm$ 0.1   & 5.7 $\pm$ 0.2 & 10.6 $\pm$ 0.2 & $0.50 \pm 0.03$ & 0.82\\
\hline
Ti:GeO$_2$-based DSD & Split & 0.7 $\pm$ 0.1 & 8.8 $\pm$ 0.2 & 11.2 $\pm$ 0.2 & $0.44 \pm 0.03$ & 0.81 \\  
\end{tabular}
\end{ruledtabular}
\label{tab:results}
\end{table*}

{\it Discussion and Outlook$-$} 
This work presents the first successful experimental realization of  multi-material optical coatings designed via multi-objective optimization,
 a milestone underscored by the flawless fabrication of two distinct prototypes.
The good agreement between prediction and measurement for both distinct and complex ternary systems proves the robustness and predictive power of our design framework. 
This establishes a reliable methodology for creating bespoke optical coatings with tailored properties that were previously inaccessible.

The realization of the SiN$_x$-based coating serves as a crucial proof-of-concept, since 
it demonstrates that even materials with relatively high absorption can be incorporated into a low-noise design.
The subsequent realization of the Ti:GeO$_2$-based coating, which leverages a more advanced but comparatively less-studied material, showcases the true potential of the validated method. 

{\it Acknowledgements $-$} The CTN experiment is supported by the LIGO Laboratory. LIGO was constructed by the California Institute of Technology and Massachusetts 
Institute of Technology with funding from the National Science Foundation and operates under cooperative agreement PHY-1764464. 
CTN data analysis and figure generation were carried out using MATLAB software provided by The MathWorks, Inc.

The authors V.P., I.M.P., V.G. gratefully acknowledge support from the Istituto Nazionale di Fisica Nucleare (INFN) via the VIRGO and VIRGO-ET projects. 
We are also grateful to the members of the LIGO Optics Working Group and the Virgo Coating Research and Development for valuable discussions.

In the document repositories of the LIGO and Virgo Collaborations, this work has numbers LIGO-P2500607-v1 and VIR-0888A-25, respectively.


\onecolumngrid    
\clearpage


\onecolumngrid    
\clearpage        

\setcounter{equation}{0}
\setcounter{figure}{0}
\setcounter{table}{0}
\renewcommand{\theequation}{S\arabic{equation}}
\renewcommand{\thefigure}{S\arabic{figure}}
\renewcommand{\thetable}{S\arabic{table}}

\begin{center}
    \textbf{\Large Supplementary Material: Experimental Realization of Optimized Ternary Mirror Coatings}
\end{center}

\vspace{0.5cm}

\begin{center}
        \textbf{Abstract} 
\end{center}

\begin{center}
    \begin{minipage}{0.85\textwidth}
        \small 
        \parskip=0.5em 

        This supplementary material details the multi-objective optimization framework used to design advanced low-noise ternary mirror coatings for next-generation gravitational wave detectors. 
The primary goal is the simultaneous minimization of coating thermal noise, optical transmittance, and absorbance. 
By employing the Borg Multi-Objective Evolutionary Algorithm heuristic, this research explores the Pareto front of optimal trade-offs, providing the theoretical foundation required for experimental implementation.

The investigation presents simulations for two promising material systems: silicon nitride/silica (SiN$_{x}$/SiO$_{2}$) and titania-doped germania/silica (Ti::GeO$_{2}$/SiO$_{2}$). 
All simulations are grounded in experimentally-derived optical and mechanical parameters from prior characterizations of single-layer materials. 
The analysis focuses on advanced ternary structures, specifically the Double Stack of Doublets design, to overcome the performance limitations of traditional binary stacks.

The results underscore the critical importance of process consistency. 
While a simulated hybrid coating, using two different deposition facilities, exhibited a fragmented solution space, assuming a single high-quality deposition process for both the SiN$_{x}$ and Ti::GeO$_{2}$-based 
systems produced smooth, well-defined Pareto fronts. 
This analysis identifies concrete, optimized Double Stack of Doublets structures for both material systems, establishing them as viable and compelling candidates for experimental realization. 
As a primary example of the achieved performance, the Ti::GeO$_{2}$-based design is predicted to yield a thermal noise reduction quantified by an Amplitude Spectral Density Reduction Factor of $0.71$,
 while meeting the stringent optical requirements of $5.6$~ppm transmittance and $0.5$~ppm absorbance.

A concluding tolerance analysis, performed via Monte Carlo simulations, confirms the robustness of the proposed designs against state-of-the-art manufacturing imprecisions. 
This work therefore provides the essential theoretical blueprint for the successful fabrication and testing of these ultra-low-noise mirrors.

    \end{minipage}
\end{center}

\vspace{0.8cm} 


\section{Methodology: Optimization Framework and Physical Models}

In this section we first introduce the multi-objective optimization framework that forms the core of our design strategy, 
followed by a description of the established physical models used to evaluate the coating's performance characteristics.
This foundation is essential for the systematic exploration and identification of novel coating designs that push beyond current performance boundaries.

\subsection{Multi-Objective Optimization Framework}
\label{sec:Ott}

In the design of high-performance dielectric mirrors for gravitational wave (GW) detectors, the engineering goal is not to optimize a single performance metric, but rather to find an 
optimal balance between two physically competing requirements. As detailed by Refs. \cite{BORGVP,VPbinary}, the problem centers on optimizing the vector of layer thicknesses, 
$\mathbf{z} = (z_1, \dots, z_{N_L})$, to simultaneously minimize the mirror's thermal noise, its optical transmittance and absorbance at the laser operating frequency ({\it single-band optimization}).

{This focus on a {\it single-band optimization} is a deliberate strategic choice. Our primary goal was to identify the absolute best design possible for the main operating wavelength (1064\,nm)
 in terms of minimal thermal noise. Introducing an additional constraint at another wavelength would inevitably restrict the available solution space and degrade the ultimate performance achievable.
For the same reason, since all objectives are calculated at this single wavelength, material dispersion was not included in the model.
However, the code supports the inclusion of dispersion in the calculation of optical properties.
}

A classical approach to such a problem would be to formulate it as a constrained optimization. 
For instance, one could seek to minimize the coating's thermal noise function, $\Phi_c(\mathbf{z})$, while ensuring that the transmittance, $\tau_c(\mathbf{z})$,  and  absorbance 
$\alpha_c(\mathbf{z})$,  remain below some pre-defined thresholds. This is mathematically expressed as:
\begin{equation}
\min_{\mathbf{z} \in \Omega} \Phi_c(\mathbf{z}) \quad \text{subject to} \quad \tau_c(\mathbf{z}) \leq \tau_0, \quad \alpha_c(\mathbf{z}) \leq \alpha_0.
\label{eq:min_noise}
\end{equation}
Here $\Omega$ defines the search space of physically allowable layer thicknesses.

A more comprehensive and powerful framework, which is used in our study, treats the problem from a true multi-objective standpoint.
{For this purpose, we employ the state-of-the-art Borg Multi-Objective Evolutionary Algorithm (Borg MOEA) heuristic \cite{BORG}. 
The choice of this specific algorithm is deliberate and offers several key advantages over traditional methods:}

\begin{itemize}
    \item {\textbf{Avoidance of Subjective Weighting:} A significant benefit of the Borg MOEA is that it eliminates the need for a single weighted cost function to account for competing objectives, as done in Ref. \cite{Venugopalan24}. 
     Such weighting schemes necessarily introduce a subjective pre-selection of constraints and can lead to an incomplete exploration of the solution space (see \cite{Marler2004}). 
     Our approach circumvents this by treating all objectives as equally important during the optimization phase.}
    \item {\textbf{Robust Pareto Front Discovery:} The Borg MOEA is specifically designed to effectively discover the complete set of objectively optimal solutions, i.e. the Pareto surface. 
    The algorithm's output is therefore the Pareto front itself—a property of the problem space, not a product of subjective preferences introduced via weights.}
    \item {\textbf{Convergence Diagnostics:} The qualitative features of the computed Pareto front, such as its smoothness and continuity, provide valuable heuristic information.
     A well-defined, continuous front is a strong indicator that the optimization process has effectively converged toward the true optimal trade-off surface, 
     giving us additional confidence in the robustness and quality of the final selected design.}
\end{itemize}

In this paradigm, the goal is to discover the entire set of all possible optimal trade-offs. The problem is thus formulated as the simultaneous minimization of an objective vector:

\begin{equation}
\min_{\mathbf{z} \in \Omega} \left[ \Phi_c(\mathbf{z}), \tau_c(\mathbf{z}), \alpha_c(\mathbf{z}) \right].
\label{eq:multi_obj}
\end{equation}

Solving this problem yields not a single solution but a set of non-dominated points forming the Pareto front, 
where improving one objective (e.g., reducing thermal noise) necessarily results in the degradation of another (e.g., increasing transmittance).
This approach provides designers with a complete map of optimal trade-offs, enabling an informed decision based on the holistic requirements of the detector, rather than being limited by an arbitrarily chosen initial constraint.
The papers \cite{BORGVP,VPbinary} demonstrate that formulations (\ref{eq:min_noise}) and (\ref{eq:multi_obj}) are mathematically equivalent, as they converge to the same solution.
However, while formulation (\ref{eq:min_noise})  follows a constrained optimization approach, formulation ~(\ref{eq:multi_obj}) adopts a true multi-objective strategy, leading to substantially different efficiency and exploration properties. As shown in our previous publications, the multi-objective formulation is not only more efficient but also enables a more exhaustive mapping of the optimal trade-offs. Consequently, exploring the Pareto front generated by formulation ~(\ref{eq:multi_obj}) with a specialized, state-of-the-art heuristic is essential for fully understanding the performance limits of the system.

To evaluate the objective functions $\tau_c(\mathbf{z})$ and $\alpha_c(\mathbf{z})$ for any given thickness vector $\mathbf{z}$, a well-established optical model is employed, commonly known as the \textit{transfer matrix method} \cite{Abele}. 

This model was implemented within our proprietary codebase to ensure a tight and efficient integration with the Borg MOEA optimization engine,
 a critical requirement for handling the vast number of evaluations inherent to the heuristic search. For this reason, we did not use external packages like \cite{Rumpf11, Luce22}.
The mirror coating is modeled as a stack of $N_L$ dielectric layers, where each layer $m$ is characterized by its thickness $z_m$ and its complex refractive index, $n^{(m)}$. 
The complex nature of the refractive index accounts for material absorption.

The thermal noise objective, $\Phi_c$, is directly related to the coating's Brownian thermal noise (see next section). 
Specifically, it is defined as the \textit{amplitude spectral density} of this noise, which is the square root of the power spectral density ($S_{\text{CB}}$),
 evaluated at a representative frequency for GW detectors, typically $f=100$ Hz. 
 For practical comparison and to create a dimensionless figure of merit, it is convenient to normalize this value by the noise produced by the mirror reference design currently in use.
This ratio will be referred to as the Amplitude Spectral Density Reduction Factor. A value less than unity indicates a performance improvement over the reference design.

A relevant special case for these mirrors is the double stack of doublets (DSD) design. 
For this specific DSD case, the optimization variables are therefore: the number of doublets $N_1$ for the first stack, along with the thickness vectors $\mathbf{z}_1$ and $\mathbf{z}_2$ for its constituent layers (each of length $N_1$); 
and the number of doublets $N_2$ for the second stack, with its corresponding thickness vectors $\mathbf{s}_1$ and $\mathbf{s}_2$ (each of length $N_2$).


The selection of the final design for fabrication is a distinct, \textit{a posteriori} step performed on the computed Pareto front. 
This is a deterministic, two-stage methodology designed to yield a unique, optimal solution:
\begin{enumerate}
    \item {First, we filter the entire front to isolate the subset of solutions that satisfies the hard operational constraints of the detector (e.g., $\tau_c(\mathbf{z}) \le \tau_0$ and $\alpha_c(\mathbf{z}) \le \alpha_0$).}
    \item {Second, from this feasible subset, we select the single design corresponding to the point with the \textbf{absolute minimum Brownian thermal noise}.}
\end{enumerate}
This two-stage process—first exploring the entire space of optimal trade-offs and then selecting based on hard constraints— therefore ensures
 the identification of a single optimal design for fabrication, guided by physical constraints rather than arbitrary cost functions.


\subsection{Modeling Coating Thermal Noise Objective Functions}

A foundational approach for describing the coating Brownian noise ($S_{\text{CB}}$) is the \textit{Braginsky model} \cite{Braginsky2002, GV}. This framework, discussed in the main letter, provides a direct calculation of thermal noise based on the materials' mechanical properties and layer thicknesses. 

For a comprehensive description, the \textit{Fejer Effective Medium Theory} \cite{Fejer2021LIGO} offers a more advanced treatment of the subject.
The key advantage of the Fejer theory is its ability to explicitly distinguish between two fundamental dissipation mechanisms:
 \textit{bulk loss} ($\phi_{\perp}$) and \textit{shear loss} ($\phi_{\parallel}$). Despite its greater complexity, this model also allows the final thermal noise to be expressed in a way that is dependent on a linear combination of layer thicknesses, 
 making it compatible with similar optimization strategies.

A significant practical challenge in applying the Fejer model is the experimental difficulty of independently measuring the bulk and shear loss angles \cite{BeS1,BeS2}. To simplify its application, 
it is common to make the approximation that these two loss components are equal ($\phi_{\perp} \approx \phi_{\parallel}$). 
Under this assumption, the Fejer model reduces to the Braginsky model, effectively framing the latter as a specific case of the more general theory.

Alternative models for describing coating thermal noise have been proposed~\cite{GV, Hong}, which are potentially more accurate.
 However, these more complex models often fail to significantly outperform simpler ones when fitting experimental data. 
 This is largely because current experimental uncertainties mask the finer details of the phenomenon.

\section{Theoretical Framework for Binary Stacks} 

The performance of a standard quarter-wavelength (QWL) binary stack is critically limited by optical absorption, especially for demanding applications like GW detectors. 
This fundamental constraint is quantified by the {\it Koppelmann limit} (henceforth KL)  \cite{Koppelmann}, which establishes the minimum possible absorption for a given pair of materials in a QWL stack.
 An approximate analytical form for the KL is given by:
\begin{equation}
    \alpha_\infty \approx \frac{2\pi}{\lambda_0} \frac{n_L \kappa_H + n_H \kappa_L}{n_H^2 - n_L^2},
    \label{eq:koppelmann}
\end{equation}
where $n$ and $\kappa$ are the refractive index and extinction coefficient for the high-index (H) and low-index (L) materials, respectively, at the operating wavelength of $\lambda_0=1064$\,nm.

To overcome the KL, Carniglia proposed a {\it tapered} design, where the layer thicknesses are apodized to minimize the electric field intensity within the more absorptive material \cite{carniglia}. 
This methodology allows the coating's absorption to approach the {\it Carniglia limit} (henceforth CL), a lower bound set purely by the bulk properties of the constituent materials. 
However, for applications as sensitive as GW detectors, this reduction in absorption often comes at the cost of increased Coating Thermal Noise (CTN), a critical performance-limiting factor.
To address this inherent trade-off, we utilize the Borg Multi-Objective Evolutionary Algorithm (Borg-MOEA) framework (see Section~\ref{sec:Ott}) for a multi-objective optimization of the mirror coatings, balancing these competing requirements.

All results presented in this work are computed using a proprietary code that implements, for optical simulations,  the exact Abel\'{e}s transfer matrix method \cite{Abele}. 
The code's accuracy is benchmarked against analytical solutions, including the limit described in Eq.~(\ref{eq:koppelmann}), which is valid for low-contrast QWL stacks.

\section{Simulation and Optimization of Advanced Coatings}
\label{sec:design_characterization}

In this section optical and mechanical parameters (refractive index, absorption,
and internal friction) of the individual materials used in the design algorithm were sourced from prior, dedicated characterizations of
single-layer samples \cite{granata20_review,granata20,Granata21} .

The purpose of the following analysis is twofold. For the Ti::GeO$_{2}$ system, our goal is to define the optimal theoretical performance, providing a clear and ambitious target for experimental validation. 
For the SiN$_{x}$ system, despite its higher intrinsic absorption, the objective is to explore whether advanced DSD optimization can nonetheless yield a competitive design, offering a valuable alternative pathway. 
This dual investigation provides the essential theoretical blueprints to guide and assess the subsequent experimental fabrication of both coating designs.

\subsection{Baseline Performance of Binary Stacks}

Our investigation is structured to first analyze the fundamental performance of these materials in 
traditional binary QWL stacks. This initial step allows us to evaluate their intrinsic absorption 
properties with respect to the KL and CL.
Subsequently, we will explore more advanced architectures to identify design strategies that can mitigate the 
critical trade-off between optical absorption and CTN.

\begin{figure}[p]
  \centering
  \includegraphics[width=0.8\textwidth]{"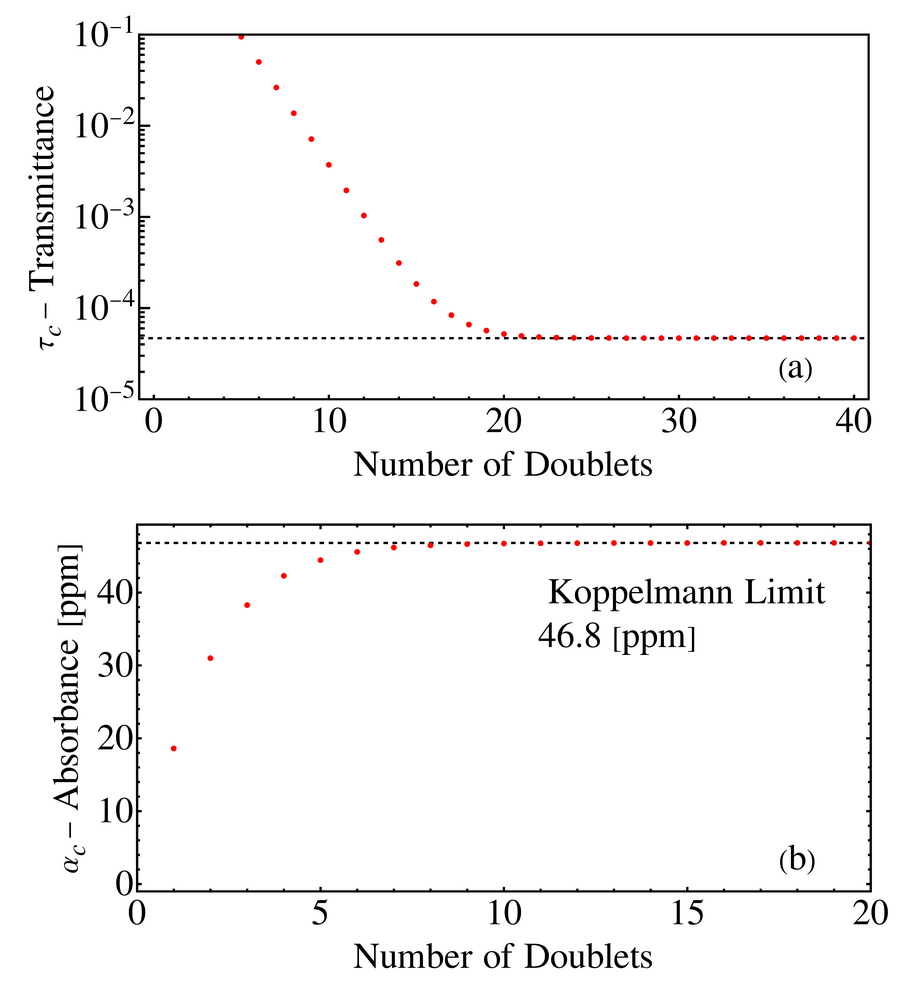"}
  \caption{(a) Transmittance ($\tau_c$) and (b) absorbance ($\alpha_c$) for a QWL multilayer stack of Silica/SiN$_x$ as a function of the number of doublets. 
  The transmittance is shown on a logarithmic scale and decreases as more layers are added.
   The absorbance, measured in parts per million (ppm), increases. 
   Both saturate at the Koppelmann Limit. The limit is $44.9$~ppm for a stack using all GC silica, and $46.8$\, ppm (as shown in this plot) when using silica of the {\it R\&D} deposition system for the bottom layers, corresponding to the hybrid design.
   Here the extinction coefficient is $\kappa_{SiN_x} = 1.5 \times 10^{-5}$.}
\label{fig:fig1}
\end{figure}

\begin{figure}[htbp]
  \centering
  \includegraphics[width=0.8\textwidth]{"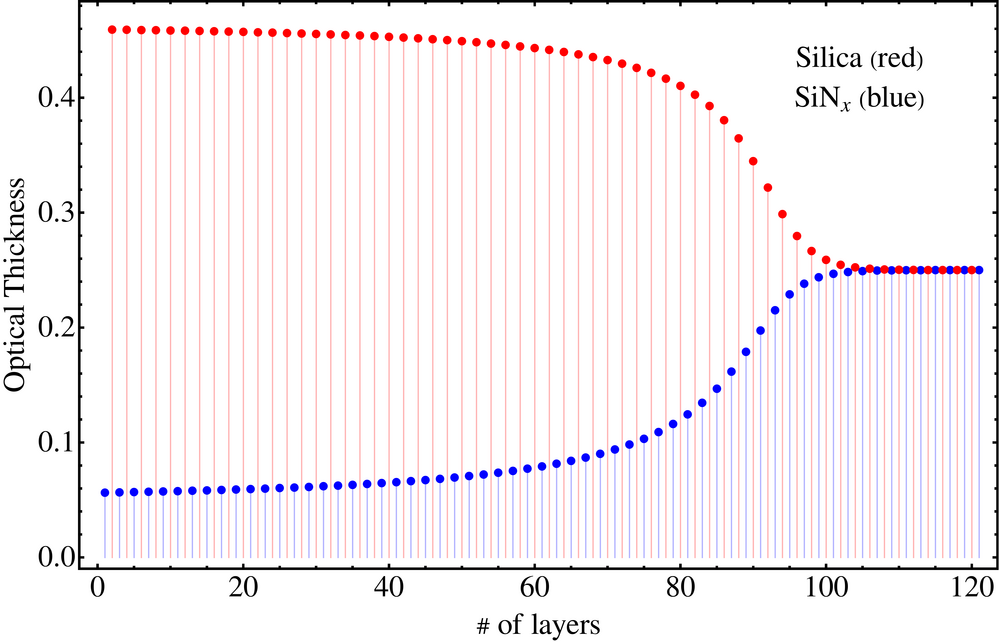"}
  \caption{Schematic representation of the tapered layer thicknesses in the SiN$_x$/SiO$_2$ stack designed according to the Carniglia method. 
  The high-index SiN$_x$ layers (blue) are progressively thinned towards the front of the mirror (left), while the low-index SiO$_2$ layers (red) are thickened to minimize absorption.}
  \label{fig:fig2}
\end{figure}

\begin{figure}[htbp]
  \centering
  \includegraphics[width=0.8\textwidth]{"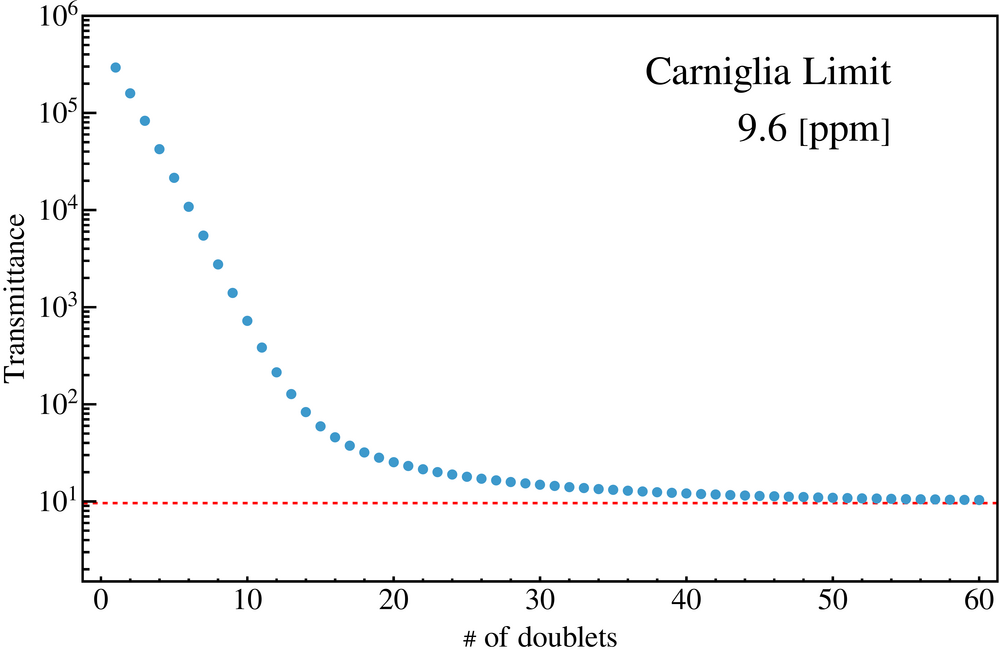"}
  \caption{Transmittance as a function of the number of layer pairs for the tapered SiN$_x$/SiO$_2$ stack. 
  The transmittance asymptotically approaches the Carniglia Limit of $9.607$ ppm, demonstrating a significant reduction in absorption compared to the standard quarter-wave stack shown in Fig.~\ref{fig:fig1}.}
  \label{fig:fig3}
\end{figure}

\begin{figure}[htbp]
  \centering
  \includegraphics[width=0.8\textwidth]{"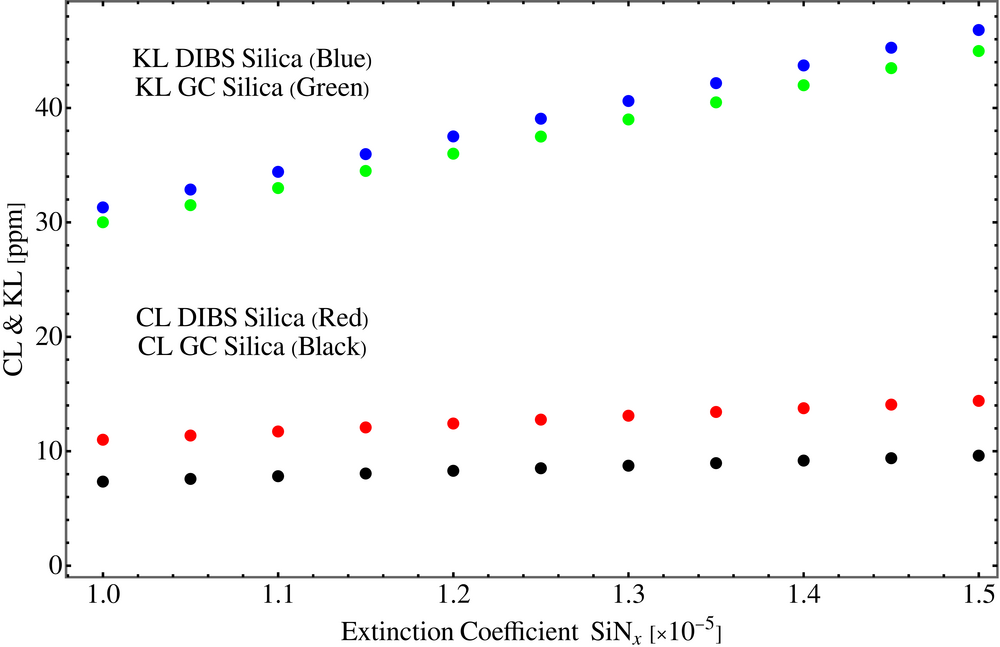"}
  \caption{Comparison of the KL and CL for SiN$_x$/SiO$_2$ stacks as a function of the SiN$_x$ extinction coefficient ($\kappa_{SiN_x}$).
   Results are shown for both GC and {\it R\&D} deposition system silica. 
   The CL is consistently lower than the KL, and the difference becomes more significant as the material absorption increases.}
  \label{fig:fig4}
\end{figure}

As a first case study, we analyze the SiN$_x$/SiO$_2$ binary system. 
Although SiN$_x$ is known to have a significant extinction coefficient \cite{LMASiNx}, its analysis is fundamental for our work. 
It serves as a critical benchmark for the predictive methodology. 
By establishing theoretical performance limits—a KL of $44.9$~ppm for a QWL stack (Fig.~\ref{fig:fig1}) and a CL of $9.607$~ppm for a tapered design (Figs.~\ref{fig:fig2}-\ref{fig:fig3})—we create 
clear quantitative targets.
Figure~\ref{fig:fig4} displays a parametric analysis of these limits as a function of extinction coefficient material properties, 
providing further details for a quantitative understanding of the model.

\begin{figure}[htbp]
  \centering
  \includegraphics[width=0.8\textwidth]{"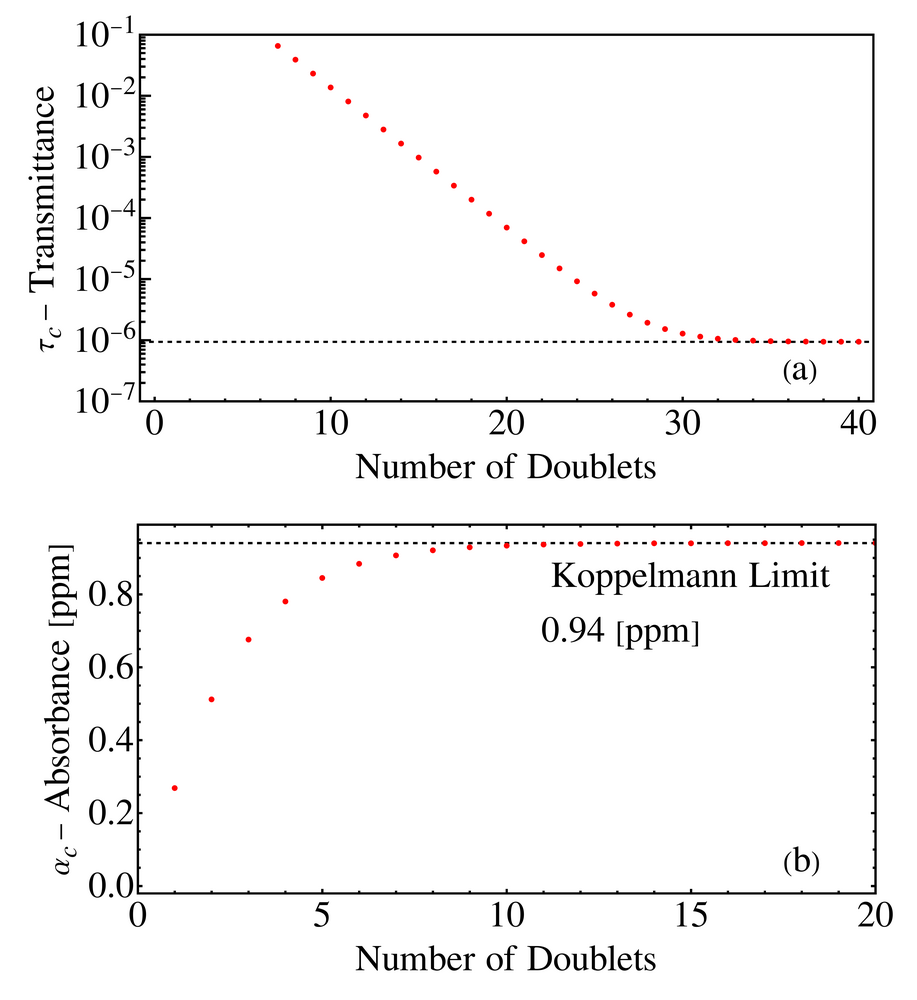"}
  \caption{(a) Transmittance ($\tau_c$) and (b) absorbance ($\alpha_c$) for a QWL multilayer stack of Silica/Ti::GeO$_2$ as a function of the number of doublets. 
  The transmittance is shown on a logarithmic scale, and the absorbance is measured in ppm. Both saturate at the Koppelmann Limit of $0.94$~ppm.}
  \label{fig:fig5}
\end{figure}

\begin{figure}[htbp]
  \centering
  \includegraphics[width=0.8\textwidth]{"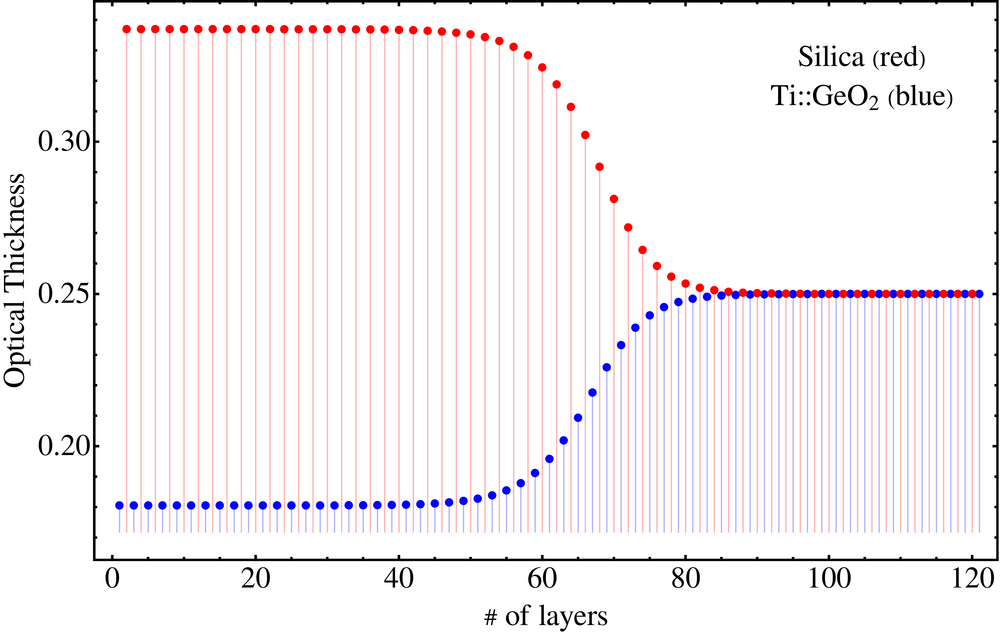"}
  \caption{Schematic of the tapered layer design for the Ti::GeO$_2$/SiO$_2$ stack, following the Carniglia optimization method to minimize absorption.}
  \label{fig:fig6}
\end{figure}

\begin{figure}[htbp]
  \centering
  \includegraphics[width=0.8\textwidth]{"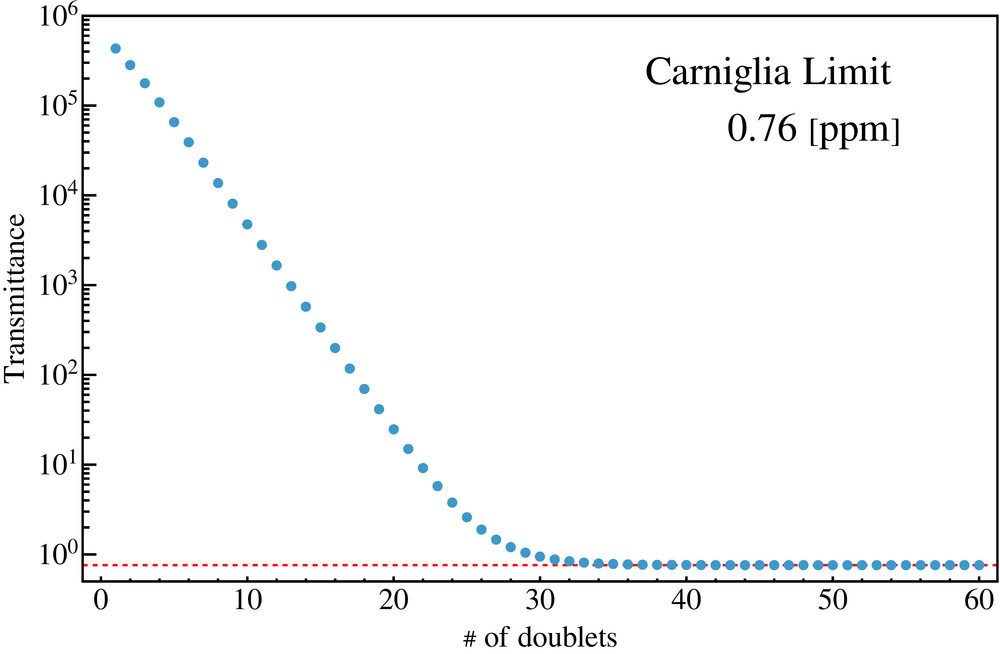"}
  \caption{Transmittance approaching the Carniglia Limit for the optimized Ti::GeO$_2$/SiO$_2$ stack. 
  The asymptotic value represents the lowest possible absorption achievable with this material pair and design methodology (i.e. $0.76$ \, ppm).}
  \label{fig:fig7}
\end{figure}

\begin{figure}[htbp]
  \centering
  \includegraphics[width=0.8\textwidth]{"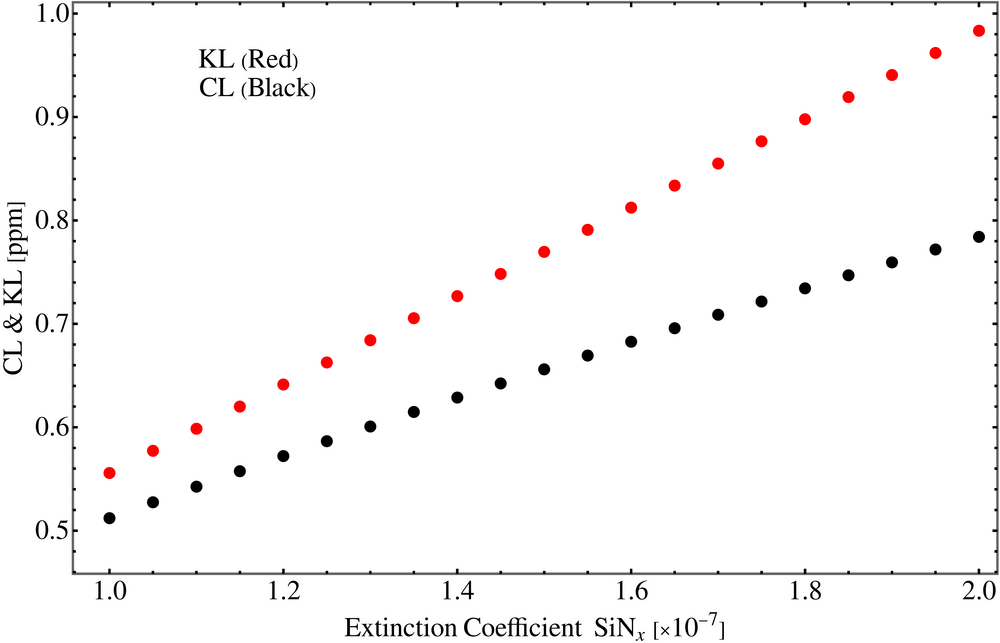"}
  \caption{Comparison of KL and CL for the Ti::GeO$_2$/SiO$_2$ system. The plot illustrates the theoretical minimum absorption for both standard (KL) and optimized (CL) designs. 
  The gap between KL and CL is smaller than for the SiN$_x$ case due to the lower extinction coefficient contrast.}
  \label{fig:fig8}
\end{figure}

The same framework is then applied to the Ti::GeO$_2$/SiO$_2$ system, a more recent material combination with lower intrinsic losses. For this stack, the calculated KL is $0.94$~ppm (Fig.~\ref{fig:fig5}) , 
which can be reduced to a CL of $0.76$~ppm through design tapering (see analysis in Figs.~\ref{fig:fig6}-\ref{fig:fig7}). 
The parametric analysis shown in Fig.~\ref{fig:fig8} reveals a crucial design principle: the potential for improvement via tapering, 
quantified by the gap between the KL and CL, is directly governed by the contrast in complex refractive index coefficients between the high- and low-index materials. For the Ti::GeO$_2$/SiO$_2$ system, 
this contrast is significantly smaller than for the SiN$_x$/SiO$_2$ stack, which explains why the gap between its KL and CL is much less pronounced.

In conclusion, this analysis reveals a fundamental performance ceiling for these binary stacks. 
While design optimization via tapering proves effective (however, increasing mechanical losses), even the high-performance 
Ti::GeO$_2$/SiO$_2$ system struggles to achieve the sub-1-ppm absorption levels required for next-generation GW detectors. 
This limitation motivates the subsequent investigation of advanced ternary architectures.
The goal will be to explore designs capable of further suppressing optical absorption without simultaneously increasing the equally critical CTN, thus breaking the performance trade-off inherent in simpler designs.

\subsection{Design of Ternary DSD Architectures}

To overcome the performance limits of binary stacks, we now investigate advanced ternary designs using a multi-objective optimization approach, in particular we use the DSD configuration.
A top stack of low-loss Ti::Ta$_2$O$_5$/SiO$_2$ was used in all cases to minimize absorption, 
while the bottom stack was chosen to be either SiN$_x$-based or Ti::GeO$_2$-based.
We employ the Borg MOEA algorithm to navigate the complex trade-off space between three competing objectives: minimizing optical absorption ($\alpha_c$), minimizing CTN 
(i.e. reduce ASD RF), and meeting the target transmittance ($\tau_c$).

This analysis focuses on the optimization landscape itself, comparing three distinct cases: an ideal silicon nitride (SiN$_x$)-based coating, assuming all deposition is performed 
by a Grand Coater (GC); a practical hybrid SiN$_x$ coating obtained in a two-step deposition process involving the {\it R\&D} deposition system followed by GC; 
and a titanium-doped germanium oxide (Ti::GeO$_2$)-based coating, also assuming all-GC deposition.

\subsubsection{Ideal SiN$_x$ Design (All-GC)}

\begin{figure}[htbp]
  \centering
  \includegraphics[width=0.8\textwidth]{"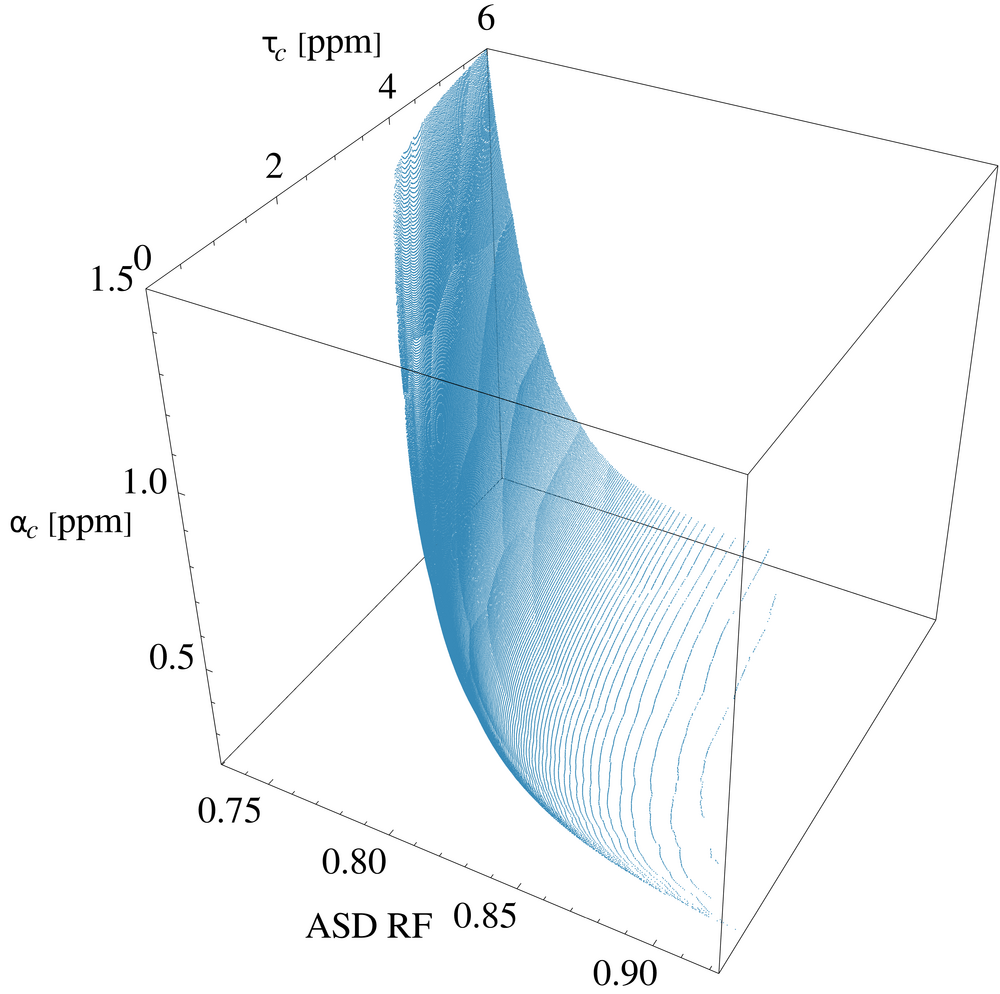"}
  \caption{Three-dimensional Pareto front for the SiN$_x$ DSD coating obtained from multi-objective optimization using Borg MOEA algorithm, 
  showing the trade-offs between optical absorption ($\alpha_c$), transmittance ($\tau_c$), and ASD RF for coating design. 
  The surface represents non-dominated solutions optimizing noise reduction performance relative to the reference coating.}
  \label{fig:fig9}
\end{figure}

\begin{figure}[htbp]
  \centering
  \includegraphics[width=0.6\textwidth]{"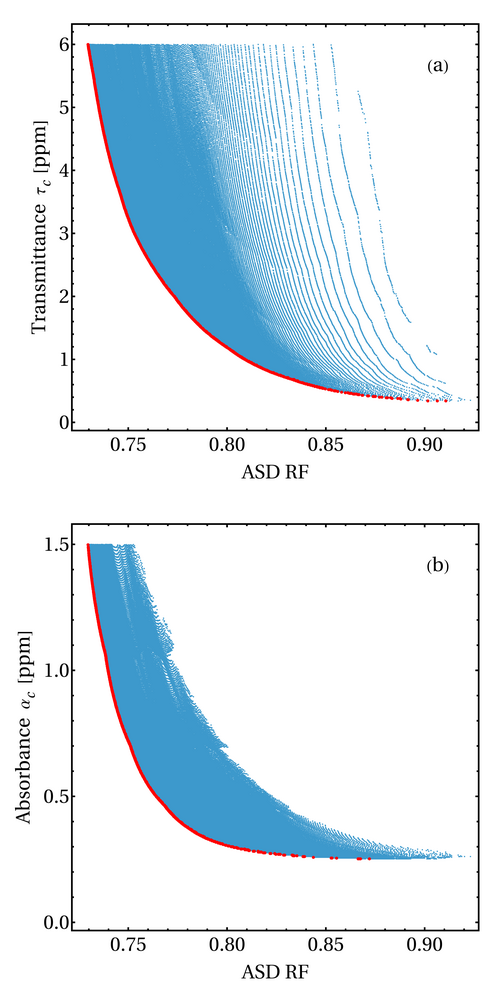"}
  \caption{Two-dimensional projections of the Pareto front for the SiN$_x$-based DSD design, illustrating the trade-offs between key performance metrics. (a) Transmittance ($\tau_c$) versus ASD RF.
   (b) Absorbance ($\alpha_c$) versus ASD RF. The red line in each plot marks the boundary of optimal solutions, representing the best achievable performance for each pair of objectives. 
   This solution space is used to select the final coating design based on performance constraints.}
  \label{fig:fig10}
\end{figure}

First, we consider an idealized SiN$_x$-based design, assuming all layers can be deposited in the high-quality GC facility. The resulting three-dimensional Pareto front, shown in Fig.~\ref{fig:fig9}, is smooth and continuous. 
This indicates a well-behaved optimization problem, where a clear and unbroken set of optimal trade-off solutions exists. 
The two-dimensional projections (Fig.~\ref{fig:fig10}) allow for a clearer visualization of the direct trade-offs between pairs of objectives, from which a final design can be selected. 
Figure~\ref{fig:fig11} shows the layer structure of one such optimal solution, uniquely determined by the constraints $\tau_c \le 5.6$\,ppm and $\alpha_c \le 1.5$\,ppm. 
This result demonstrates a successful outcome of the optimization process for this ideal case.

\subsubsection{Hybrid SiN$_x$ Design}

We then analyze a more practical scenario where the coating is produced via a hybrid sputtering process. 
In this method, an initial SiN$_x$-based stack is deposited on the substrate using the {\it R\&D} system, after which a second stack of Silica/Titania-doped Tantala is deposited on top via GC sputtering.
This case contrasts sharply with the ideal one. 
The resulting Pareto front (Fig.~\ref{fig:fig12}) is fragmented and discontinuous. 
This is a critical finding: the use of higher-loss {\it R\&D} system-deposited silica in the bottom part of the stack changes the structure of the solutions space. 
This distinct behavior of the  {\it R\&D} system-silica renders the system {\it effectively quaternary}, a condition for which the standard ternary DSD optimization strategy is ill-suited.

This fragmentation severely constrains the available solution space, creating {\it forbidden} regions of performance, particularly visible in the projection between absorbance and thermal noise (Fig.~\ref{fig:fig13}). 
The scattered nature of the optimal solutions highlights the performance degradation and the difficulty in finding a robust design.
Figure~1a of the main letter shows the DSD structure of the minimal noise design 
selected from the Pareto front, which satisfies the constraints 
$\tau_c \le 5.6~\text{ppm}$ and $\alpha_c \le 1.5~\text{ppm}$. However, due to the aforementioned fragmentation and discontinuity of the 
Pareto front, this design achieves an absorbance value of 
$\alpha_c = 1.44~\text{ppm}$.

\subsubsection{Ti::GeO$_2$ Design (All-GC)}

Finally, we apply the optimization framework to the Ti::GeO$_2$-based system, where all layers are deposited using the consistent, high-quality GC process. 
As shown in Fig.~\ref{fig:fig17}, the resulting Pareto front is once again smooth and continuous, analogous to the ideal SiN$_x$ case. 
This reinforces the conclusion that process consistency and high-quality materials are key to a well-behaved optimization landscape. 
The 2D projections in Fig.~\ref{fig:fig18} illustrate the clear trade-offs, and Fig.~1b,  displayed in the letter,
 presents the optimal layer recipe selected from this well-defined front.

\begin{figure}[htbp]
  \centering
  \includegraphics[width=0.8\textwidth]{"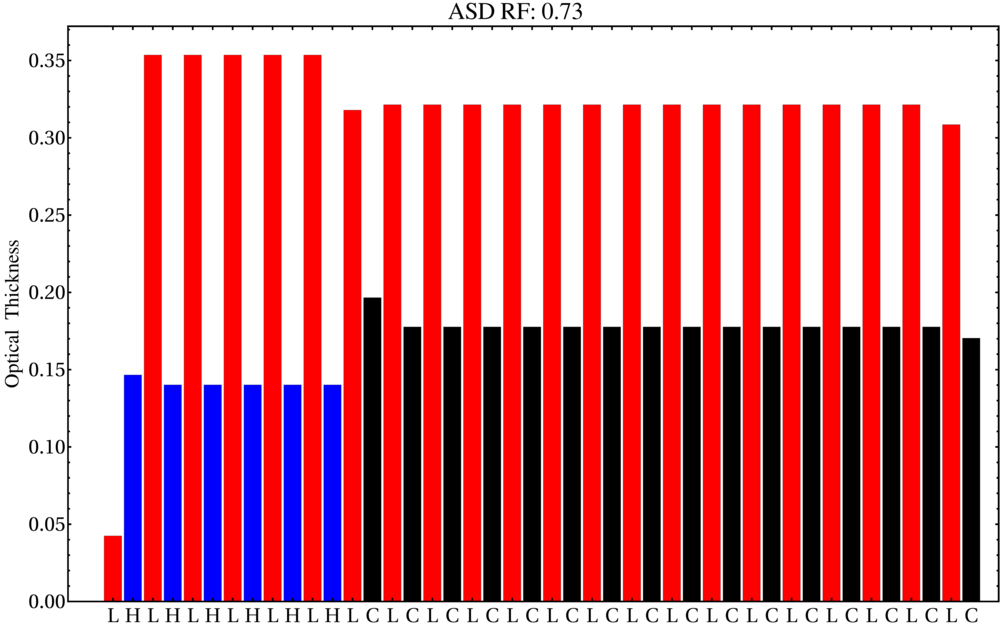"}
  \caption{The optimal layer structure selected from the Pareto front for the ideal SiN$_x$ design. 
  This design corresponds to the solution in the top-left region of the plots in Fig.~\ref{fig:fig10},
   minimizing thermal noise while meeting the optical constraints. 
   The bar chart details the optical thickness of each layer in the stack, the structure is (LH)$^6$(LC)$^{16}$ (here $N_{1}=6$ and $N_{2}=16$), where the labels L, H, and C denote silica, titania-doped tantala, and SiN$_x$ respectively.}
  \label{fig:fig11}
\end{figure}

\begin{figure}[htbp]
  \centering
  \includegraphics[width=0.8\textwidth]{"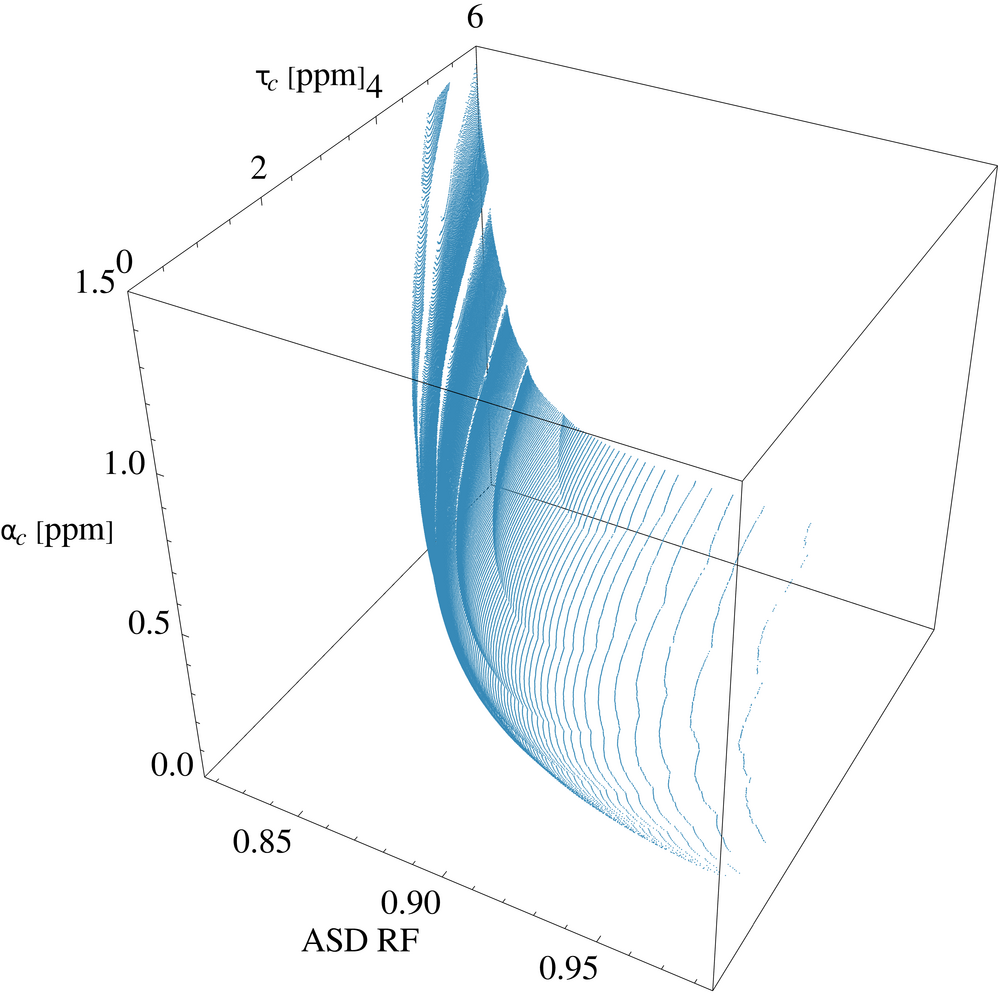"}
  \caption{
  Three-dimensional Pareto front for the hybrid SiN$_x$ based DSD coating prepared by {\it R\&D} system-GC. 
  The discontinuity in the front highlights a constrained solution space and forbidden performance regions, which are attributed to the higher optical loss of the {\it R\&D} deposition system silica. 
  This distinct behavior of silica renders the system effectively quaternary, indicating that the DSD optimization strategy is ill-suited for this material combination.}
  \label{fig:fig12}
\end{figure}

\begin{figure}[htbp]
  \centering
  \includegraphics[width=0.6\textwidth]{"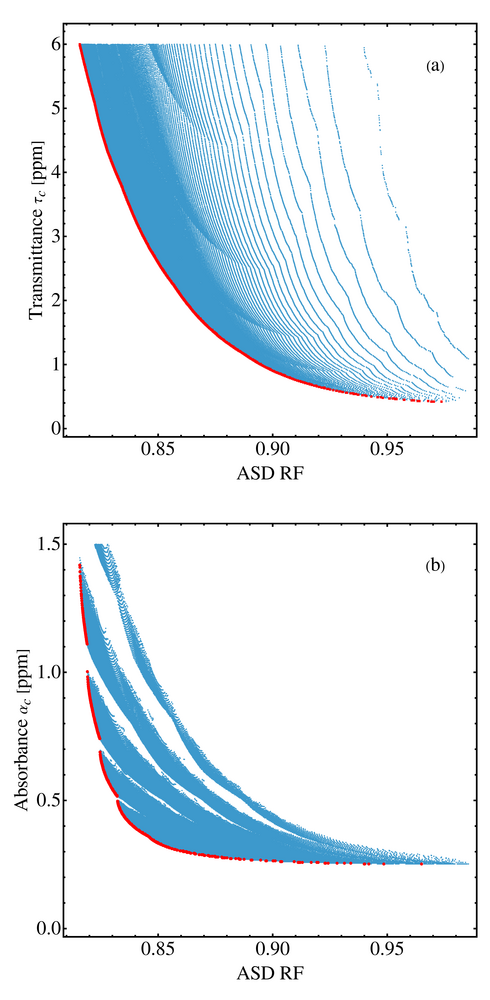"}
  \caption{Two-dimensional projections of the Pareto front for the hybrid SiN$_x$-based DSD design. 
  The fragmentation and scattering of the optimal solutions, especially when compared to Fig.~\ref{fig:fig10}, highlight the performance degradation caused by the use of higher-loss {\it R\&D} deposition system silica in the bottom stack.
  }
  \label{fig:fig13}
\end{figure}

In conclusion, from the set of Pareto-optimal solutions, we selected the design that minimizes the ASD RF while satisfying the optical constraints ($\tau_c \le 5.6$\,ppm and $\alpha_c \le 1.5$\,ppm for SiN$_x$, $\alpha_c \le 0.5$\,ppm for Ti::GeO$_2$). 
These designs, for the three cases, are shown in Fig.~\ref{fig:fig11}, and in Fig. 1a and b of the letter.

\subsection{Comparative Performance and Manufacturing Tolerance}

\begin{figure}[htbp]
  \centering
  \includegraphics[width=0.8\textwidth]{"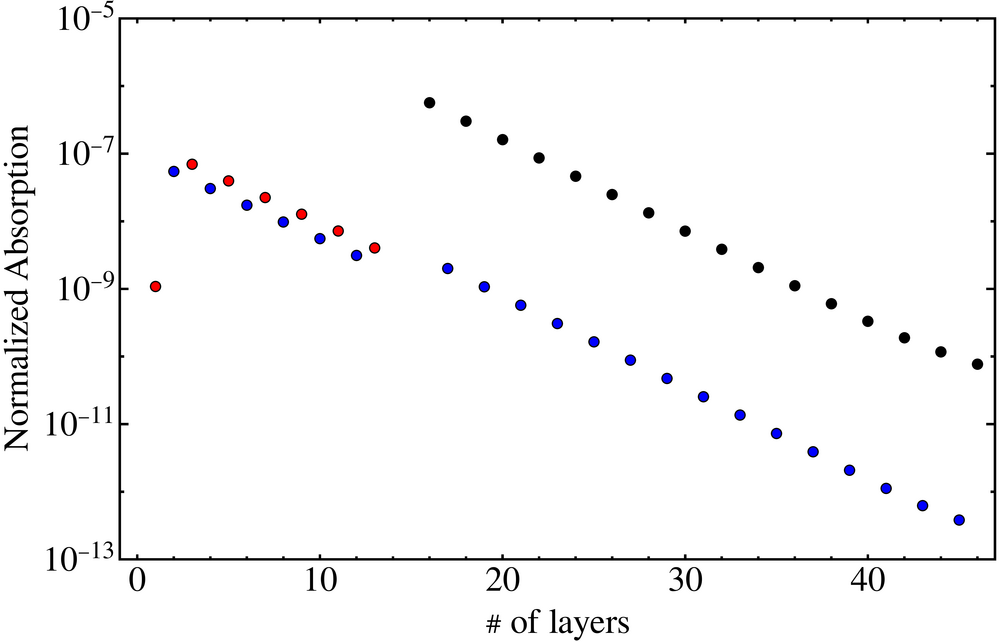"}
  \caption{Layer-by-layer contribution to the total absorption for the ideal SiN$_x$ design. Total absorbance of the coating is $1.5$\,ppm.
  The colors identify the material for each layer: SiN$_x$ (black), SiO$_2$ (red), named L, and Ti::Ta$_2$O$_5$ (blue). SiN$_x$ is the dominant contributor.}
  \label{fig:fig16}
\end{figure}

\begin{figure}[htbp]
  \centering
  \includegraphics[width=0.8\textwidth]{"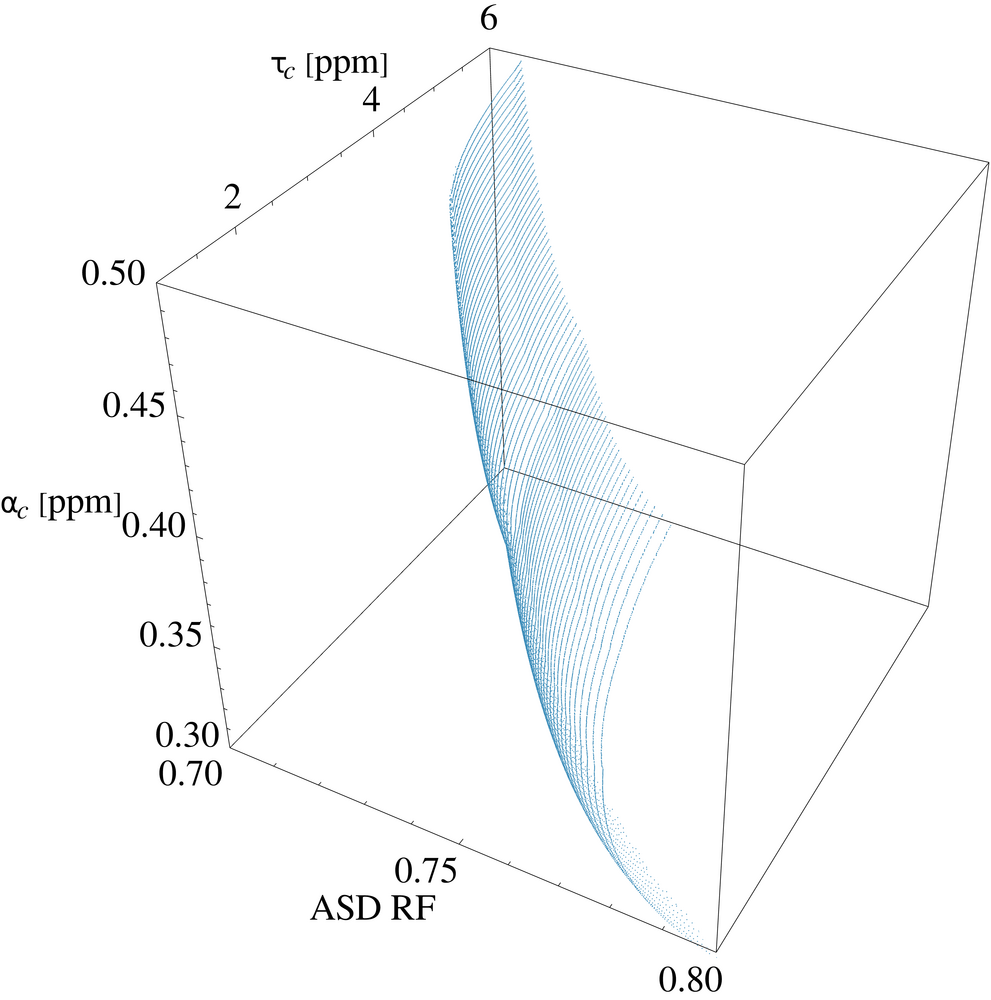"}
  \caption{Three-dimensional Pareto front for the Ti::GeO$_2$-based DSD coating. 
  The smooth, continuous surface illustrates the well-defined trade-offs achievable with an all-GC deposition process.}
  \label{fig:fig17}
\end{figure}

\begin{figure}[htbp]
  \centering
  \includegraphics[width=0.6\textwidth]{"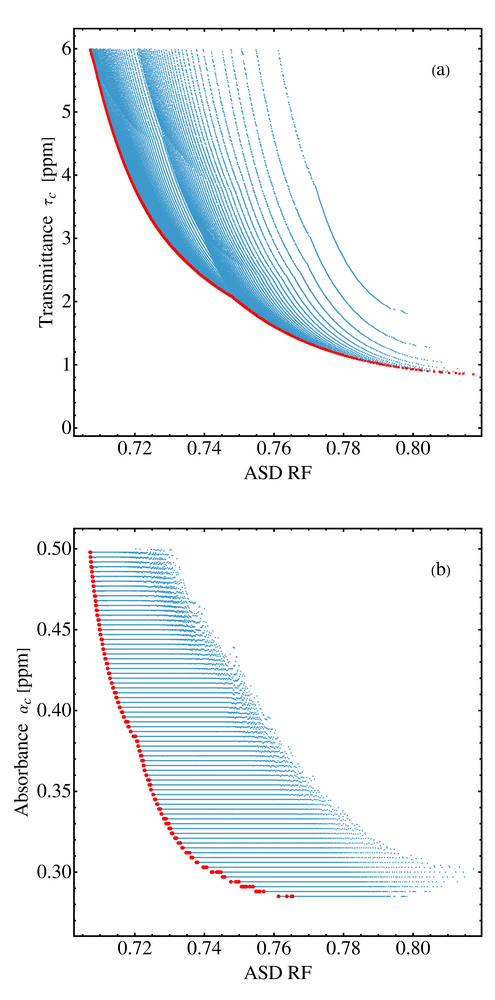"}
  \caption{Two-dimensional projections of the Pareto front for the Ti::GeO$_2$-based DSD design. \textbf{(a)} Transmittance ($\tau_c$) versus ASD RF.
   \textbf{(b)} Absorbance ($\alpha_c$) versus ASD RF. The solid red lines mark the boundary of optimal solutions.}
  \label{fig:fig18}
\end{figure}

\begin{figure}[htbp]
  \centering
  \includegraphics[width=0.8\textwidth]{"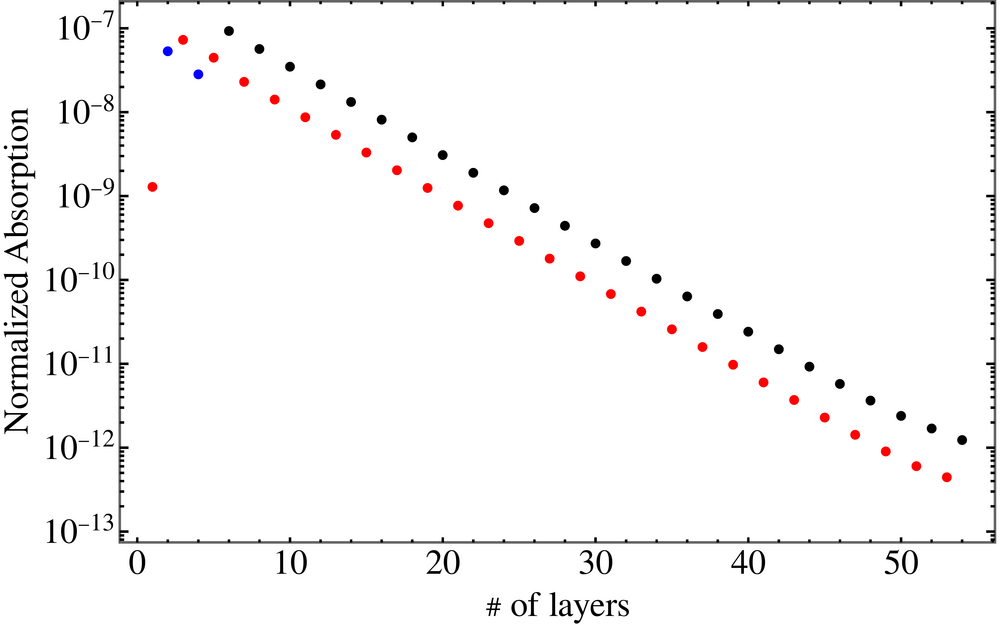"}
  \caption{Layer-by-layer contribution to total absorption for the optimized Ti::GeO$_2$-based design. The total absorbance is $0.5$\,ppm. 
  The colors identify the material for each layer: Ti::GeO$_2$ (black), SiO$_2$ (red), and Ti::Ta$_2$O$_5$ (blue). 
  The Ti::GeO$_2$ layers are the main, but manageable, contributors to absorption.}
  \label{fig:fig21}
\end{figure}

A comparative analysis of the optical properties reinforces the above conclusions. The transmittance spectra for all optimal designs meet the high-reflectivity requirement at $\lambda_0=1064$\, nm  as displayed in Figs.~3a and b of the letter.
However, the layer-by-layer absorption profiles reveal the core limitation. For the SiN$_x$ design, the SiN$_x$ layers are the dominant source of absorption, with contributions orders of magnitude higher than any other material (Fig.~\ref{fig:fig16}). 
This makes it fundamentally challenging to reach sub-ppm levels. 
In contrast, although the Ti::GeO$_2$ layers are also the primary absorbers in their stack, their intrinsic optical loss is low enough that the overall coating absorption remains at the $0.5$\,ppm target (Fig.~\ref{fig:fig21}). 
This highlights the success of the DSD architecture in managing absorption (i.e. it is possible to {\it circumvent} the CL) by placing the lossier  materials deeper in the stack, where the electric field is weaker.


{To ensure the practical viability of the optimized coating designs, a tolerance analysis was performed to assess their robustness against manufacturing errors. 
We performed a  Monte Carlo simulation ($10^5$ trials) that  incorporates random perturbations on both the layer thicknesses and the refractive indices of the materials. 
The methodology and parameters for this expanded analysis are detailed below:}

\begin{itemize}
    \item {\textbf{Perturbed Parameters:} The simulation introduces random, independent uncertainties to two key parameters. 
    Firstly, the physical thickness of each layer, as this is the quantity directly controlled and subject to error 
    during the deposition process. Secondly, the refractive index for each material.}
    
    \item {\textbf{Uncertainty Magnitudes:} The independent thickness uncertainty of $\pm 0.5$\,nm
    (modeled as a uniform distribution) is representative of the state-of-the-art precision achievable with the GC \cite{LMA}, that is the
     advanced ion-beam sputtering systems used for fabricating mirrors for gravitational-wave detectors. The uncertainty for the refractive indices was modeled based on typical run-to-run variations observed in our 
     deposition systems, corresponding to a relative uncertainty of 0.2\%.}
\end{itemize}

\begin{figure}[htbp]
  \centering
  \includegraphics[width=0.8\textwidth]{"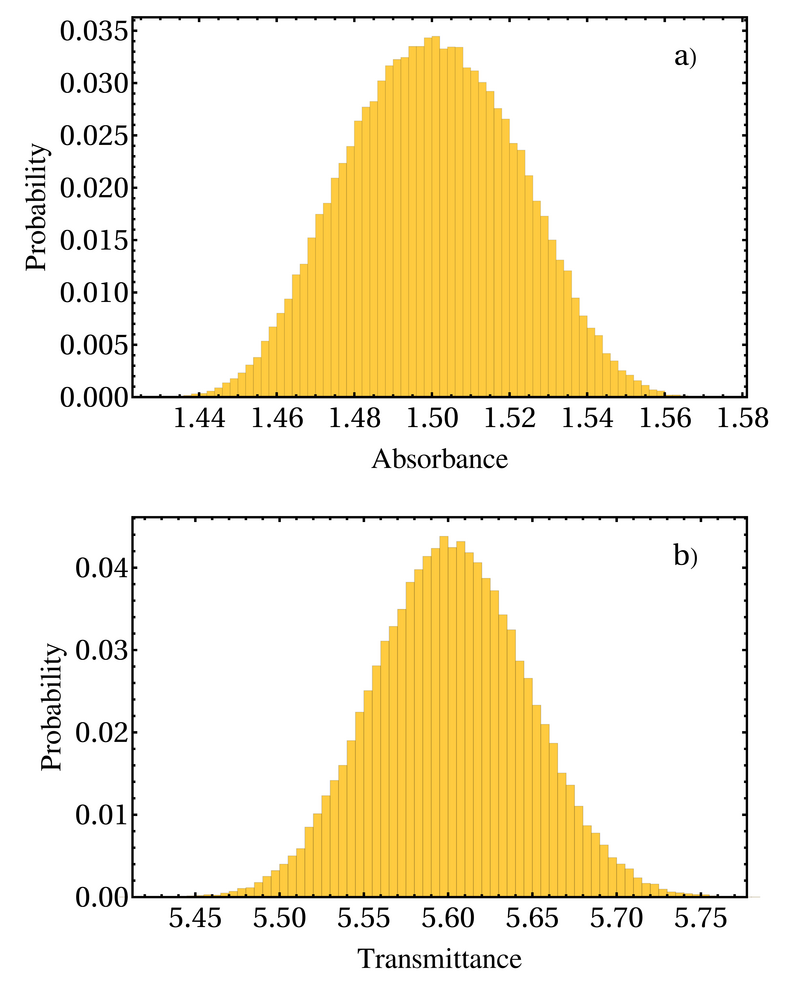"}
  \caption{Tolerance analysis for the optimized SiN$_x$-based design. 
  The plots show the statistical distribution of (a) absorbance and (b) transmittance resulting from a Monte Carlo simulation ($10^5$ trials) 
  where an independent random thickness uncertainty of $\pm 0.5$\, nm per layer and 0.2\% uncertainty on refractive index. 
  The results show values centered at $1.5$ ppm for absorbance and $5.6$ ppm for transmittance (that are the nominal values).}
  \label{fig:fig22}
\end{figure}

\begin{figure}[htbp]
  \centering
  \includegraphics[width=0.8\textwidth]{"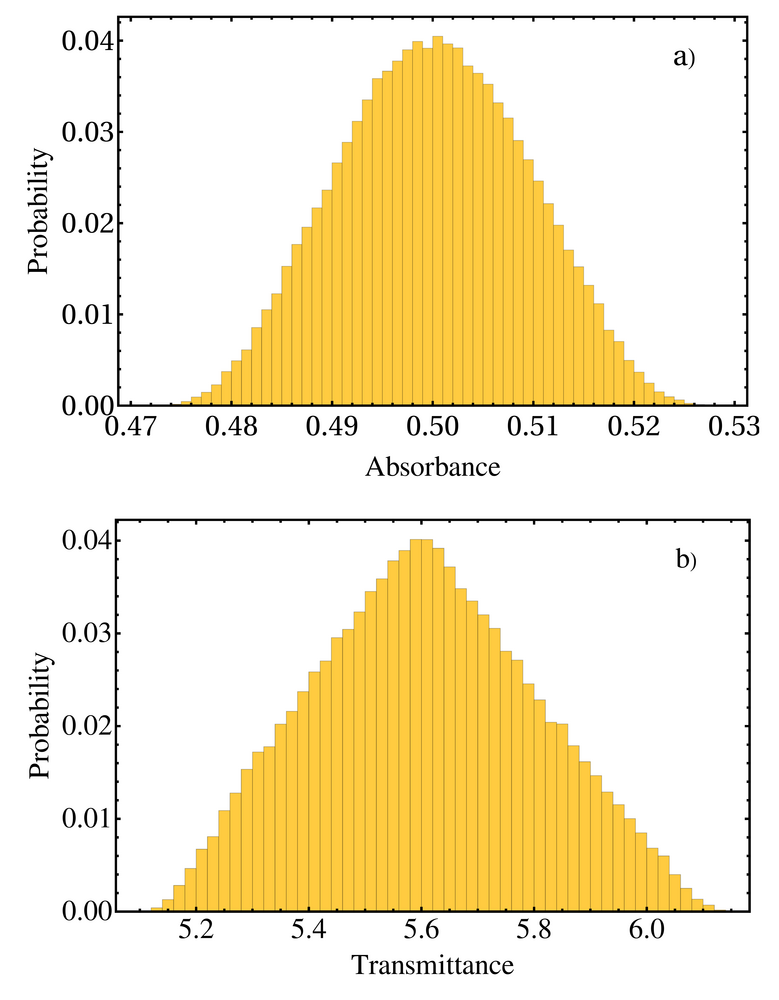"}
  \caption{Tolerance analysis for the optimized Ti::GeO$_2$-based design. 
  The plots show the statistical distribution of (a) absorbance and (b) transmittance resulting from a Monte Carlo simulation ($10^5$ trials) where an independent random thickness error of $\pm 0.5$\, nm 
  was applied to each layer {and 0.2\% uncertainty on refractive index}. 
  The narrow distributions are centered at the target values of $0.5$ ppm  for absorbance and $5.6$ ppm for transmittance (that are the nominal values).
}
\label{fig:fig23}
\end{figure}

The resulting statistical distributions for the key performance metrics of absorbance and transmittance are shown in the Figs. \ref{fig:fig22}, and \ref{fig:fig23}.
The analysis reveals that the designs are highly robust. The narrow probability distributions for both absorbance (a) and transmittance (b) indicate that the optical properties 
are stable and not overly sensitive to typical random errors in layer thickness and refractive index. 
Furthermore, when subjected to the same random thickness errors, the resulting ASD RF values form a distribution that is approximately Gaussian.
This distribution is exceptionally narrow, with a standard deviation on the order of 0.01\%  relative to the mean value for both the proposed design.

A summary of the final performance metrics is presented in Table~\ref{tab:final_summary}. 
The results show that  Ti::GeO$_2$-based DSD design  
achieves the lowest thermal noise (ASD RF = $0.71$) while simultaneously meeting the aggressive sub-$0.5$\, ppm absorbance target.


\begin{table}[!htbp]
    \centering
    \caption{Summary of predicted performance. This table compares the fundamental absorption limits of the binary material pairs (KL, and CL) 
    with the final achieved metrics of the optimized ternary DSD designs. 
    The results highlight how the DSD architecture allows the final coating to achieve performance levels beyond the binary limits. All values are theoretical predictions intended as experimental targets.}
    \label{tab:final_summary}
    \begin{tabular}{|l|c|c|c|c|c|}
        \toprule \hline \hline
        Design Case & Binary KL & Binary CL &  ASD RF &  Transmittance $\tau_c$ & Absorbance $\alpha_c$ \\
         & [ppm] & [ppm] & & [ppm] & [ppm] \\
        \midrule \hline
        Ideal SiN$_x$ (all GC) & 44.9 & 9.61 & 0.73 & 5.6 & 1.50 \\
        Hybrid SiN$_x$ ({\it R\&D} deposition - GC) & 46.8 & 14.4 & 0.82 & 5.6 & 1.44 \\
        Ti::GeO$_2$ (all GC) & 0.94 & 0.76 & 0.71 & 5.6 & 0.50 \\
        \bottomrule \hline \hline
    \end{tabular}
\end{table}

\end{document}